\documentclass[namedreferences]{SolarPhysics}
\usepackage[optionalrh]{spr-sola-addons}
\usepackage{solaheader} 
\usepackage{epsfig}          
\usepackage{graphicx}        
\usepackage{color}           
\usepackage{url}             

\begin{document}

\begin{article}

\begin{opening}

\title{Interpreting Helioseismic Structure Inversion Results of Solar Active Regions}

\author{Chia-Hsien~\surname{Lin}$^{1a,2}$\sep
        Sarbani~\surname{Basu}$^{1b}$\sep
        Linghuai~\surname{Li}$^{1c}$
        }

\runningauthor{C. H. Lin {\it et al.}}
\runningtitle{Interpreting Inversion Results of Active Regions}

\institute{$^{1}$ Department of Astronomy, Yale University,
                     P.O. Box 208101, New Haven, CT 06520-8101, USA\\
                  email a: \url{linc@tcd.ie} \\
                  email b: \url{sarbani.basu@yale.edu} \\
                  email c: \url{li@astro.yale.edu} \\
           $^{2}$  School of Physics, Trinity College Dublin, 
                   Dublin 2, Ireland\\
           }

\begin{abstract}
Helioseismic techniques such as ring-diagram analysis have often  been used to determine
the subsurface structural differences between solar active and quiet regions.
Results obtained by inverting the frequency differences between the
regions  are
usually interpreted as the sound-speed differences between them.
These in turn are used as a measure of 
temperature and magnetic-field strength  differences between the
two regions. 
In this paper we first show that the ``sound-speed'' difference
obtained from inversions is actually 
a combination of sound-speed difference and a magnetic component.
Hence, the inversion result is
not directly related to the thermal structure. 
Next, using solar models that include magnetic fields,
we develop a formulation to use the inversion results 
to infer the differences in the magnetic and thermal structures between
active and quiet regions.
We then apply our technique
to existing structure inversion results for different pairs
of active and quiet regions. 
We find that
the effect of magnetic fields is strongest in a shallow region above
0.985$R_\odot$
and that
the strengths of magnetic-field effects at the surface and
in the deeper ($r < 0.98R_\odot$) layers are inversely related,
{\it i.e.}, the stronger the surface magnetic field
the smaller the magnetic effects in the deeper layers, and {\it vice versa}.
We also find that
the magnetic effects in the deeper layers are the strongest
in the quiet regions,
consistent with the fact that these are basically regions 
with weakest magnetic fields at the surface.
Because the quiet regions were selected to precede or follow 
their companion active regions, the results could have implications
about the evolution of magnetic fields under active regions.
\end{abstract}


\end{opening}

\section{Introduction}
\label{sec:intro}

The availability of high-precision helioseismic data and
the advancement of local helioseismology techniques
have enabled investigations of fine details of solar internal dynamics for small
regions of the Sun.
Some of the methods, 
such as ring-diagram and time-distance analyses,
have also enabled us examine details of
the structure below solar active regions. 

Ring-diagram analysis (Hill, 1988; Patron {\it et al.}, 1997) 
is used to determine 
frequencies 
of  short-wavelength (high degree)
modes in a small region of the Sun. These are modes that
can be approximated as plane waves over a small area of the Sun. 
These frequencies can be inverted to determine the structure of the
region under study using the same techniques that are
used to invert global modes to study the spherically symmetric part of 
the solar structure. 
To avoid systematic errors, structure
inversions are usually done using frequency differences between
two regions to determine the difference in structure between the two regions
({\it e.g.}, Basu, Antia and Bogart, 2004, 2007).
Time-distance helioseismology (Duvall {\it et al.}, 1993) 
measures the variations in the travel time
of acoustic waves. The travel-time variations can then be used to infer 
the variations in the wave speed and flow velocity 
through an inversion procedure 
({\it e.g.}, Kosovichev, Duvall, and Scherrer, 2000; 
Kosovichev, Duvall, and Birch, 2001).

Basu, Antia, and Bogart (2004) inverted the frequency differences 
between several active regions
and nearby quiet regions to determine the sound-speed difference and 
the difference
in the adiabatic index $\Gamma_1$
between these pairs of regions. They found that
for all the active region -- quiet region pairs, 
the sound speed below the active regions was
smaller than that of the quiet regions for about the first 7 Mm, 
and then the sound speed became larger in the active regions than 
in the quiet regions. A similar behavior was seen
for the adiabatic index ($\Gamma_1$). 
Qualitatively similar results were obtained 
by Kosovichev, Duvall, and Scherrer (2000) and
Kosovichev, Duvall, and Birch (2001) using time-distance analysis.
Kosovichev, Duvall, and Scherrer (2000),
Kosovichev, Duvall, and Birch (2001), and
Basu, Antia, and Bogart (2004)
interpreted the results in terms of a 
temperature difference, and alternatively as a difference in the magnetic field.
However, 
by examining the $\Gamma_1$ differences,
Basu, Antia, and Bogart (2004) also concluded that 
the results cannot be explained as being caused by temperature changes or 
magnetic fields alone.

The presence of magnetic fields can affect the frequencies of waves
in two ways:
Firstly, the magnetic fields change the thermodynamic structure
({\it i.e.}, pressure and temperature profiles) of the medium that
the waves travel through, which, in turn, changes the frequencies of the waves.
Secondly, the plasma waves are directly affected by the magnetic fields
through the Lorentz force.
In other words,
the modification to the frequencies results from
both structural and non-structural ({\it i.e.}, Lorentz force) 
effects of the magnetic fields.
Since the two effects are inseparable in the observed frequencies,
the ``structures'' revealed by the inversions could partly be
the manifestations of
the frequency difference caused by Lorentz force on wave propagation.
In addition,
the inversion kernels to-date have been derived from
non-magnetic reference models.
Since the structural variables of a magnetic model have
both magnetic and non-magnetic (gas) components,
it had been uncertain if an inversion using a non-magnetic reference model would
reveal the whole or only the gas part of the structural change.
By using solar models that include magnetic fields
and inversion kernels computed from non-magnetic reference models,
Lin, Li, and Basu (2006)
showed that the ``sound speed'' revealed by the inversions
in the presence of magnetic fields
is in fact a combination of both sound speed
($c_g \equiv \sqrt{\Gamma_1 P_{\rm gas}/\rho}$) and
Alfv{\'e}n speed ($c_A \equiv B/\sqrt{4\pi\rho}$).
To distinguish this property from the actual sound speed,
we call it ``wave speed'',
defined as $c_T \equiv \sqrt{\Gamma_1 P_T/\rho}$,
where $P_T$ is the total pressure
($=P_{\rm gas}+P_{\rm mag}$,
where $P_{\rm mag} = B^2/8\pi$, $B$ being the magnetic-field strength).
While the
speed of sound is directly related to temperature,  there is
no simple, direct relationship between the ''wave speed'' $c_T$ and
either temperature or magnetic fields.
Hence, it has not been possible to determine the
magnetic-field and temperature profiles below active regions
with any degree of accuracy.

In this paper,
we use solar models that include magnetic fields to first confirm
the Lin, Li, and Basu (2006) results and then derive
a practical way of using the 
wave speed and $\Gamma_1$ inversion results to determine the thermal and
magnetic structural differences between active and quiet regions.
After confirming the reliability of this method to infer
the thermal and magnetic structures,
we then apply it
to the inversion results of Basu, Antia, and Bogart (2004).

The main limitation of our models is that they are one dimensional, and hence
the form of magnetic fields allowed is quite restrictive. We cannot, for
example, have toroidal magnetic fields or magnetic fields over restricted
ranges of latitudes or longitudes. However, we are forced to use such models
because the codes to construct self-consistent solar models 
with arbitrary magnetic
fields do not yet exist. A consequence of using 1D models is that we are likely
to overestimate the effect of magnetic fields on thermal transport. 
In the real,
three-dimensional case, 
convection, when obstructed by a magnetic flux tube can bend around the tube, 
which is not possible in the 1D case.

The rest of the paper is
organized as follows: 
We describe the magnetic models in Section~\ref{sec:mdl}.
A brief description of the inversion technique and 
the inversion results obtained from
pairs of models are given in Section~\ref{sec:inv}. 
The strategy to link the inversion results and 
temperature and magnetic fields is derived in Section~\ref{sec:calib}. 
We apply the strategy to solar data in Section~\ref{sec:sun}, 
and discuss the results in Section~\ref{sec:interpretation}.
The result of an attempt to model the subsurface magnetic structure is 
presented in Section~\ref{sec:model}.
We summarize our results in Section~\ref{sec:summary}.
It should be noted that although we apply our method to results
from ring-diagram analyses, it can be applied to any helioseismic
technique that can determine the adiabatic index ($\Gamma_1$) below active 
regions.

\section{Models}
\label{sec:mdl}

Since ring-diagram analysis is a variant of global-mode analysis,
we model the difference between quiet and active regions as the difference
between a solar model with no magnetic fields and
a solar model that contains magnetic fields and the associated effects.
We use a modified version of YREC (the Yale Rotation and Evolution Code:
Demarque {\it et al.}, 2008) to construct the models.

How magnetic effects are incorporated in the models has been described in detail
by Li and Sofia (2001) and Li {\it et al.}, (2003). 
We give a brief overview here.

To compute the effects of the magnetic fields, 
two magnetic variables are introduced.
They are the magnetic energy density ($\chi \equiv B^2/8 \pi \rho$)
which describes the magnetic perturbation strength,
and 
the magnetic field direction
[$\gamma \equiv (2B_t^2 + B_p^2)/B^2$]
which accounts for the tensorial feature of the magnetic pressure.
$B_t$ and $B_p$ are respectively the toroidal and poloidal components of the
magnetic field. 
The magnetic pressure can then be expressed in terms of $\chi$, $\gamma$, and
$\rho$, that is, $P_{\rm mag}=(\gamma-1)\chi\rho=B_t^2/8\pi$,
in which the fact that
the pressure is a tensor in the presence of magnetic fields
has been taken into account.

With the addition of the magnetic variables,
the equations of magnetic models are as following
(Lydon and Sofia, 1995; Li and Sofia, 2001; Li {\it et al.}, 2003):

\begin{eqnarray}
\frac{\partial P_T}{\partial M_r} &=&
- \frac{GM_r}{4\pi r^4} - \frac{1}{4\pi r^2}\frac{\partial^2 r}{\partial t^2},
\label{eq:eom} \\
\frac{\partial r}{\partial M_r} &=&
\frac{1}{4\pi r^2 \rho},
\label{eq:mass} \\
\frac{\partial L}{\partial M_r} &=&
\epsilon - T \frac{dS_T}{dt} - \frac{1}{\rho}\frac{\partial u}{\partial t},
\label{eq:energy} \\
\frac{\partial T}{\partial M_r} &=&
- \frac{T}{P_T} \frac{G M_r}{4\pi r^4} \nabla,
\label{eq:trans_conv}
\end{eqnarray}
and the first law of thermodynamics becomes:
\begin{equation}
T{\rm d}S_T=  {\rm d}Q_T = {\rm d}U_T + P_T{\rm d}V - 
                          (\gamma - 1)(\chi/V){\rm d}V.
\label{eq:second}
\end{equation}
This modification thus affects
the luminosity equation
and
the energy-transport equations
(Equation~\ref{eq:energy} and Equation~\ref{eq:trans_conv}).
In the above equations, $M_r$ is the mass inside a radius $r$,
$\epsilon$ is the energy generation rate,
$u = a T^4$ is the radiation energy density ($a$ being radiation constant),
and $\nabla$ is the  dimensionless
temperature gradient when magnetic fields are included.

These equations are similar to those governing non-magnetic models
except that
the pressure, total internal energy, and entropy now have a magnetic term,
that is: $P_T = P_{\rm gas} + P_{\rm mag}$, 
$U_T = U + \chi$ and $S_T = S + \chi/T$. 
The solar models are constructed with OPAL opacities 
(Iglesias and Rogers, 1996) at high
temperatures and Alexander and Ferguson (1994) opacities at low temperatures.
Nuclear reaction rates
of Bahcall and Pinsonneault (1992) were used. 
It should be noted that since we are
dealing with the outermost layers of the model, the nuclear reaction rates only play
an indirect role.
The equation of state (EOS) of the gaseous phase is from OPAL 
(Rogers and Nayfonov, 2002).
However, this
has to be modified in the presence of magnetic fields.
The equation of state is formally
$\rho=\rho(P_T, T, \chi, \gamma)$, 
in addition to the usual dependence on chemical abundances:
\begin{equation}
\frac{{\rm d}\rho}{\rho} = \alpha \frac{{\rm d}P_T}{P_T} - 
\delta \frac{{\rm d}T}{T} - 
\nu \frac{{\rm d}\chi}{\chi} - \nu' \frac{{\rm d}\gamma}{\gamma},
\label{eq:eos}
\end{equation}
plus terms due to composition.
In the equation, $\alpha$, $\delta$, $\nu$, and $\nu'$ 
are partial derivatives with respect to
$P_T$, $T$, $\chi$ and $\gamma$, respectively.

Thus Equation~\ref{eq:second} can be re-written as:
\begin{equation}
  T{\rm d}S_T=C_P{\rm d}T-\frac{\delta}{\rho}{\rm d}P_T+\left(1+
        \frac{P_T\delta\nu}{\rho\alpha\chi}\right){\rm d}\chi+
        \frac{P_T\delta\nu'}{\rho\alpha\gamma}{\rm d}\gamma.
\end{equation}

The adiabatic index $\Gamma_1$ can be written in the same usual form, 
{\it i.e.},
$\Gamma_1=1/(\alpha-\delta\nabla'_{\rm ad})$, 
where $\nabla'_{\rm ad}$ is the modified
adiabatic gradient and is given by
\begin{equation}
\nabla'_{\rm ad}=\nabla_{\rm ad}\left( 1-{\nu\nabla_\chi\over\alpha}
                                -{\nu'\nabla_\gamma\over\alpha}\right),
\end{equation}
where, 
as usual, 
$\nabla_{\rm ad}=\partial \ln T / \partial \ln P_T=P_T\delta/\rho C_pT$,
$\nabla_\chi = \partial \ln \chi /\partial\ln P_T$ and
$\nabla_\gamma=\partial\ln\gamma/\partial\ln P_T$.
The derivatives $\alpha$, $\delta$, and $C_p$
change in the presence of magnetic fields.

The way that the EOS has been implemented is as follows: 
we determine ``uncorrected'' density
[$\rho_0=\rho_0(P_T,T)$] at a given chemical composition 
using the OPAL equation of state tables.
The density is then corrected for the missing effects 
to obtain the actual density 
[$\rho=\rho_0/(1+\rho_0 \chi(\gamma - 1)/P_T)$].
We then calculate the changes in  $\alpha$, $\delta$, and $C_p$
caused by the presence of magnetic fields,
and then use those to calculate $\Gamma_1$.
The changes in $\alpha$, $\delta$, and $C_p$ also change energy transport.
We  use the
mixing-length approximation, but with $\nabla_{\rm ad}$ replaced by
$\nabla'_{\rm ad}$ and with the modified values of
$\alpha$, $\delta$, $C_p$, {\it etc}.
This modifies the properties of convection.
There are also changes in the properties of convection.
The details of including the magnetic effects in models
can be found in
Lydon and Sofia (1995), Li and Sofia (2001)
and Li {\it et al.} (2003).

The magnetic profile of our models is specified by
the distribution of $\chi$.
We have chosen a Gaussian profile for $\chi$
because
a Gaussian profile has an unambiguous peak,
its sphere of influence is well defined, and it has no discontinuities
that can create havoc with both stellar structure and subsequent frequency
calculations.
Furthermore,
other complex fields can be constructed by simply using several Gaussian profiles.

The Gaussian profile of $\chi$ is of the following form:
\begin{equation}
\chi = \chi_0 \exp[- \frac{1}{2} (M_D - M_{Dc})^2 / \sigma^2].
\end{equation}
$M_D \equiv \log [1 - M_r/M_\odot]$.
$M_{Dc}$ and $\sigma$ are
the adjustable parameters for 
the depth (in terms of $M_D$)
of the peak of the magnetic field and
the width (dimensionless) of the profile, respectively.
$\chi_0 \equiv B_0^2/8\pi\rho_c$, in which 
$B_0$ is the tuning parameter for the peak magnetic field strength and
$\rho_c$ is the density at $M_{Dc}$.

In short,
different magnetic models are only distinguished by their magnetic profiles,
which are specified by three parameters:
$B_0$ (peak strength), $M_{Dc}$ (peak location) and $\sigma$ (width).

Although this simple profile is implemented in most of our models 
to study the relation between the ``wave speed'' and the magnetic and thermal structures,
we have also constructed other types of field profiles by using multiple Gaussians
in order to check the validity of the relation in a complex field.

\begin{figure}
\centerline{
\includegraphics[width=0.8\linewidth]{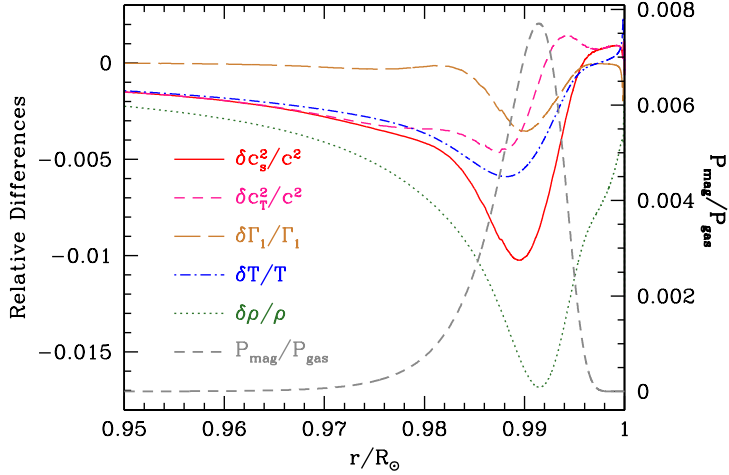}}
\caption{
Differences between a magnetic and a non-magnetic model. Note that for
the non-magnetic reference model, $c=c_T=c_g$. The gray dashed line shows the
ratio of the magnetic to gas pressure, and the value can be read from the
right-hand ordinate.
}
\label{fig:1}
\end{figure}

The relative differences of several structural parameters
between a magnetic and a non-magnetic model are shown in Figure~\ref{fig:1}.
We define $\beta \equiv P_{\rm mag}/P_{\rm gas}$.
As $P_{\rm mag} = 0$ in the non-magnetic reference model,
$\delta \beta = \beta = P_{\rm mag}/P_{\rm gas}$ illustrates the magnetic configuration
in the magnetic model in this case.
It is plotted as a gray dashed line using the scale at the right side.
We can see that $\delta c_g^2/c^2$ and $\delta c_T^2/c^2$ are very different
($c=c_T=c_g$ in the non-magnetic reference model).
The wave-speed difference ($\delta c_T^2/c^2$)
changes from negative to positive in the region of the magnetic fields
while $\delta c_g^2/c^2$ shows a single dip in the region.
The positive and negative peaks of $\delta c_T^2/c^2$ do not show any explicit
correlation with the profile of the magnetic fields.
Unsurprisingly,
$\delta c^2_T/c^2$, $\delta c^2_g/c^2$,  and $\delta T/T$ become
almost identical
at greater distances from the region of magnetic effects,
which confirms that $\delta c_T^2/c^2$ is indeed a good representation of
temperature difference in non-magnetic regions.
Another important feature is that,
except for $\delta \Gamma_1/\Gamma_1$,
all variables
show the influence of the magnetic field at large distances
from the field. The
$\delta \Gamma_1/\Gamma_1$ profile approximately follows
the  profile of $\beta$
with no tail below the region of the magnetic fields.
In other words,
the magnetic effect on $\Gamma_1$ is fairly localized 
to the vicinity of the magnetic fields.
Although the situation may change in two-dimensional models, 
where more realistic
magnetic configurations are used,
the basic features in
the differences of the structure parameters are expected to remain the same.

\section{Test Inversions}
\label{sec:inv}
\subsection{Inversion Procedure}\label{subsec:inv_tech}

The frequencies of solar oscillation modes depend on
the solar structure.
The starting point of helioseismic  inversions  is the
linearization of the oscillation equations around a known solar model (the so-called
reference model) using the variational principle.
The frequency differences can then be related to the relative variations
in sound speed ($c$) and density ($\rho$) 
either between two models or between the
Sun and the reference model.
The  relation between the differences in frequency and
these two variables ({\it i.e.}, $c$ and $\rho$)
can be written as (Dziembowski, Pamyatnykh and Sienkiewicz, 1990; Antia and Basu, 1994):
\begin{eqnarray}
\frac{\delta \omega_i}{\omega_i} &=&
\int_0^R K^i_{c^2, \rho}(r) \frac{\delta c^2}{c^2}(r) {\rm d}r +
\int_0^R K^i_{\rho, c^2}(r)
     \frac{\delta \rho}{\rho}(r) {\rm d}r +
\frac{F_{\rm surf}(\omega_i)}{E_i} + \epsilon_i,
\label{eqn:inv}
\end{eqnarray}
where
$c$ is the adiabatic sound speed,
$K^i$ are the  kernels and are known functions of the reference model,
$\delta\omega_i/\omega_i$ is the relative frequency difference of the $i$th
mode, 
$\epsilon_i$ is the observational error in $\delta\omega_i/\omega_i$, and
$F_{\rm surf}(\omega_i)/E_i$, usually called the ``surface term'',
represents the effect
of uncertainties in the model close to the surface.
Here, $E_i$ is a measure of the mode inertia.
Other pairs of variables such as ($\Gamma_1$, $\rho$) can be used instead of 
the ($c^2$, $\rho$) pair used above.
There are several techniques that can be used to invert Equation~\ref{eqn:inv}.
The inversion results used in this paper were obtained using the
Subtractive Optimally Localized Averages (SOLA) technique
(Pijpers and Thompson, 1992, 1994).
The details of the inversion procedure
and the quality of the averaging kernels were shown in
Basu, Antia, and Bogart (2004).

Basu, Antia, and Bogart (2004) 
inverted the frequency differences between active and quiet
regions,
instead of between active regions and a solar model.
One of the reasons for doing so was to reduce  foreshortening effects.
The projection of the spherical solar surface onto a flat map introduces some
foreshortening that depends on the distance of the region from the disc center,
which can cause systematic errors in determining the mode characteristics.
By selecting comparison regions at the similar latitudes and 
tracking both in time
intervals symmetric about their central meridian passages (CMP),
the authors assured that the
foreshortening effects in both regions are nearly identical, and, hence, the
effect is eliminated to a large degree by taking the difference. 
Such a difference also reduces the so-called ``surface term'' in the inversion.
A large part of the surface term in frequency differences between solar models 
and
the Sun arises from shortcomings in modeling the near-surface layers of the Sun.
Since the featured inversions are for two sets of solar frequency differences,
the surface term is much smaller, and only one surface term suffices.
The accuracy of their inversion results were ensured by
the consistency between the
the results from SOLA and the results from
regularized least squares (RLS).
It should be noted that the inversion kernels were calculated without any
effect of the magnetic fields included, and as a result the ''sound speed''
inversions yield results that are a combination of the real sound speed
difference and a magnetic component.

To simulate
the aforementioned ring-diagram inversions,
we use a non-magnetic model as the
reference model.
However, since quiet regions may also have 
magnetic fields in the deeper regions, 
we have also examined inversions between two magnetic models
while still using non-magnetic kernels.
We constructed a number of different magnetic models as the test models in
order to
determine $\delta c^2/c^2$, $\delta\Gamma_1/\Gamma_1$, and $\delta\rho/\rho$
between the magnetic models and the non-magnetic reference model. 
In order to  estimate the errors in the  inversion results 
that are caused by the errors in the observed frequencies,
we select only those  modes 
that are also present in the observational data.
The observational data set we use for the mode selection
is from Rhodes {\it et al.} (1998),
which is based on 61 days of data collected by the {\it SOI/MDI} instruments
beginning in May 1996.
The data set contains {\it f} modes and {\it p} modes up to $\ell=1000$. 
This set
is much larger than data sets obtained from ring-diagram analyses.
However,
since our
objective is merely to determine what variables are actually obtained 
by the usual ``sound-speed'' and $\Gamma_1$ inversions,
a larger data set will not affect our conclusion. 
We have also used the
360-day mode-set from MDI (Schou, 1999).

\subsection{Results} \label{subsec:result1}
The magnetic effects on thermal structures are different at different depths.
This is not only because the ratio of magnetic to gas pressure depends
on the location, 
but also because any related changes in temperature can shift the
position of the ionization zone.
Hence,
we need to examine the inversion results at different depths
in order to investigate how the variations of magnetic profiles
affect different structural properties 
({\it i.e.}, $c_g$, $c_T$, $\Gamma_1$, and $\rho$)
and the difference between $c_g$ and $c_T$,
as well as to determine whether or not  
the inversion can consistently and accurately reveal these properties 
under different magnetic situations.

\begin{figure}
\centerline{
\includegraphics[height=0.8\linewidth]{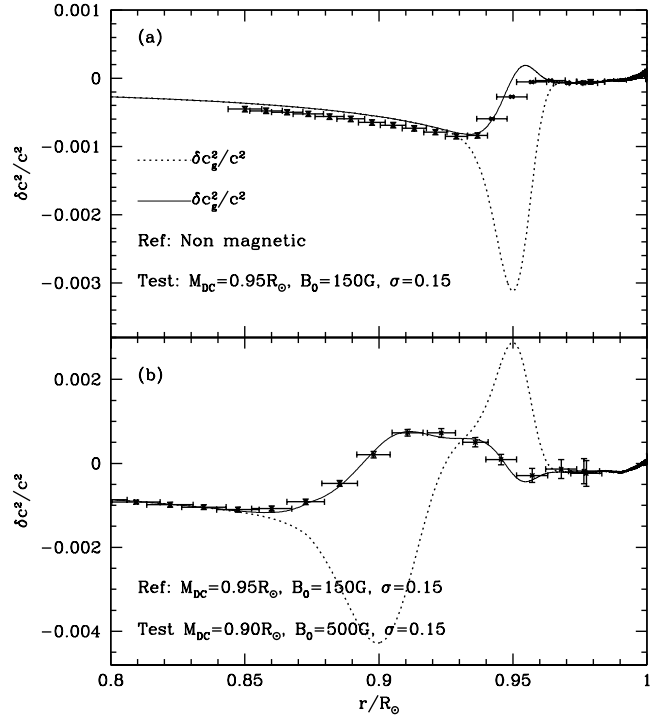}}
\caption{
Comparison of the inverted $\delta c^2/c^2$ with
the exact relative differences of the sound speed 
and of the wave speed.
The dotted line is the sound-speed difference
[$(c_g^2 - c_{\rm ref}^2)/c_{\rm ref}^2$]
and the solid line is wave speed difference
[$(c_T^2 - c_{\rm ref}^2)/c_{\rm ref}^2$],
where $c_{\rm ref}^2$ refers to $c^2$ of the reference model.
Note that $c_{\rm ref}$, $c_T$, and $c_g$ are the same
in the non-magnetic reference model.
The parameters of the magnetic-field profile
are indicated in each panel.
The inverted values are plotted as symbols along with the error bars.
The errors are obtained from the observational errors in the frequencies.
The inversion in (a) are with the Rhodes {\it et al.} mode-set, 
the one in (b) is
with the Schou (1999) set.
}
\label{fig:2}
\end{figure}

In Figure~\ref{fig:2} we show the results of ``sound-speed'' inversions
for two cases. In panel (a) we show the inversion results when one of
the models is non-magnetic, and in panel
(b) we show the results of inverting the frequency differences between
two magnetic models. 
In each case, the inversion kernels were from a non-magnetic
model. The figure shows that what the inversion produces is not the
usual ``sound speed'' difference, {\it i.e.}, difference of the quantity
$c_g^2=\Gamma_1 P_{\rm gas}/\rho$, but a ``wave speed'' difference
where the ``wave speed'' is defined as $c_T^2 = \Gamma_1 P_T/\rho$,
thereby confirming the results of Lin, Li, and Basu (2006).

\begin{figure}
\centerline{
\includegraphics[height=0.8\linewidth]{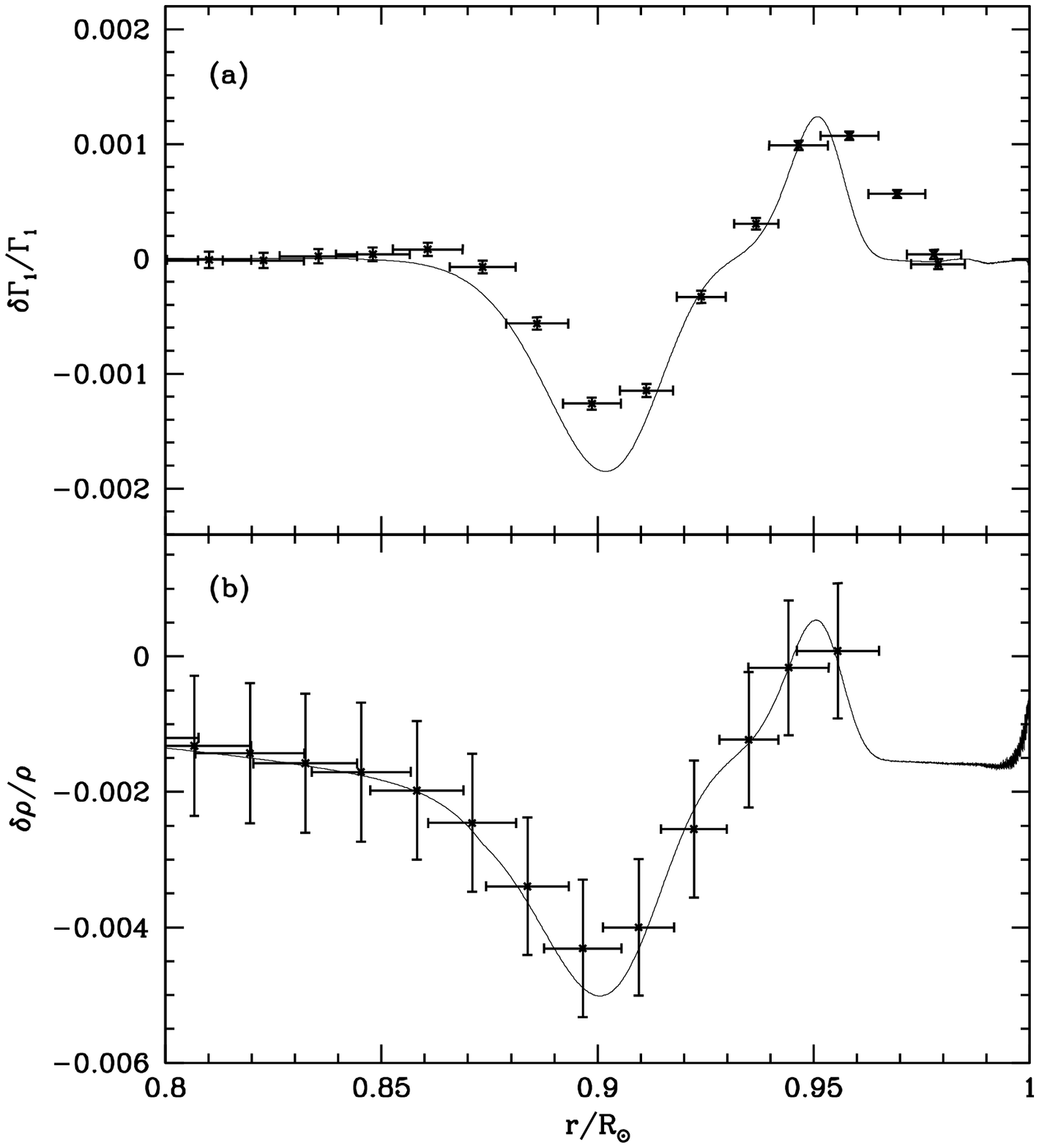}}
\caption{
Inversion results for $\delta \Gamma_1/\Gamma_1$ (upper panel) and
$\delta \rho/\rho$ (lower panel).
In both panels,
the solid line is the exact difference  between
the magnetic model and the reference model 
({\it cf.} Figure~\ref{fig:2}(b) for details of the models).
The inverted results are plotted as symbols along with the error bars.
The errors are obtained from the observational errors
of the modes used in the inversion.
The inversions are with the Schou (1999) mode set.
}
\label{fig:3}
\end{figure}

Figure~\ref{fig:3} shows the inversion results of
$\delta\Gamma_1/\Gamma_1$ [panel (a)]  and 
$\delta\rho/\rho$ [panel (b)]. In both cases we used the same models
as in Figure~\ref{fig:2}(b).
Because the adiabatic index ($\Gamma_1$) is a function of the equation of state and
because 
density ($\rho$) is an intrinsic property of gas,
it is not sensible to split them into magnetic and non-magnetic components.
$\delta\Gamma_1/\Gamma_1$ and $\delta\rho/\rho$
reflect the changes in the equations of state and in the gas distribution
caused by the presence of magnetic fields.
From Figure~\ref{fig:3},
we can see that the inversion results closely follow the computed values
even though the inversion kernels were derived from a non-magnetic model.
However, the errors in the $\delta \rho/\rho$ inversions
are too large for the results to be useful.
With the limited mode sets obtained from 
ring-diagram analysis,
the errors would be even larger,
and hence only in very rare cases
can a proper density inversion be done.


Although the frequencies of both magnetic and non-magnetic models
were computed from our pulsation code, which does not include
the Lorentz-force effect,
the effect of the absence is clearly not significant.
If it were,
the inversion results would have shown large deviations for models with
different types of magnetic profiles since
 the direct effects on wave frequencies would be very different
for different modes 
that have different lower turning points with respect to
the field positions. Clearly, such deviations are not seen.
The inversion results also show that the use of non-magnetic kernels do not
lead to any significant errors in the inversion results and hence we
are justified in using the results of Basu, Antia, and Bogart (2004).

Thus, the examples in Figures~\ref{fig:2} and \ref{fig:3} show that
the structural differences revealed 
by the inversions implemented with non-magnetic kernels
are a combination of both direct magnetic effects and
the gaseous structural changes.

\section{Estimating Magnetic and Thermal Structure from Inversion Results}
\label{sec:calib}
\subsection{Mathematical Relation between Inversion Results and Magnetic Fields}
\label{subsection:calib_math}

The wave speed ($c_T$) is affected by both the magnetic fields and the
thermal structures.
In the following, 
we derive the relations between the relative wave-speed difference 
($\delta c_T^2/c_T^2$) 
and the differences in the  magnetic structures (represented by $\delta \beta$)
and in the thermal structures 
(reflected in $\delta c_g^2/c_g^2$ or $\delta T/T - \delta \mu/\mu$).
The subscripts 1 and 0 in the equations 
denote the test and reference models wherever there may be confusion.
We should point out that
the reference model in the following derivation 
can be either magnetic or non-magnetic.
When $P_{{\rm mag}_0} = 0$,
we obtain
the special case of a non-magnetic reference model.

\begin{eqnarray}
\frac{\delta c_T^2}{c_T^2} &=&
\frac{\delta \Gamma_1}{\Gamma_1} - \frac{\delta \rho}{\rho} + \frac{\delta P_T}{P_T} \nonumber \\
&=&
\frac{\delta \Gamma_1}{\Gamma_1} - \frac{\delta \rho}{\rho} + 
\frac{\delta P_{\rm gas} + \delta P_{\rm mag}}{P_{{\rm gas}_0} + 
P_{{\rm mag}_0}} \nonumber \\
&=&
\frac{\delta \Gamma_1}{\Gamma_1} - \frac{\delta \rho}{\rho} + 
(\frac{\delta P_{\rm gas}}{P_{{\rm gas}_0}} + \frac{\delta P_{\rm mag}}{P_{{\rm gas}_0}}) 
\left[ 1 + \beta_0  \right]^{-1} ,
\label{eqn:ct}
\end{eqnarray}
where $\beta_0=P_{{\rm mag}_0}/P_{{\rm gas}_0}$ 
is the magnetic to gas pressure ratio in a reference model. 
The quantity
$\delta P_{\rm mag}/P_{{\rm gas}_0}$ can be related to 
$\delta \beta=\delta (P_{\rm mag}/P_{\rm gas})$ 
as follows:
\begin{eqnarray}
\delta \beta &=& \beta_1 - \beta_0 =
\frac{P_{{\rm mag}_1}}{P_{{\rm gas}_1}} - \frac{P_{{\rm mag}_0}}{P_{{\rm gas}_0}},
\label{eqn:beta}\\
\frac{\delta P_{\rm mag}}{P_{{\rm gas}_0}} &=&
\frac{ P_{{\rm mag}_1} - P_{{\rm mag}_0}}{P_{{\rm gas}_0}} =
\delta \beta + \beta_1 \cdot \frac{\delta P_{\rm gas}}{P_{{\rm gas}_0}}.
\label{eqn:beta_pmag}
\end{eqnarray}
By using Equation~(\ref{eqn:beta_pmag}) and assuming
$\beta_1 \ll 1$ and $\beta_0 \ll 1$,
Equation~(\ref{eqn:ct}) becomes:
\begin{eqnarray}
\frac{\delta c_T^2}{c_T^2}
&\approx&
\frac{\delta \Gamma_1}{\Gamma_1} - \frac{\delta \rho}{\rho} + \frac{\delta P_{\rm gas}}{P_{\rm gas}} +
\delta \beta (1 + \frac{\delta P_{\rm gas}}{P_{\rm gas}}) 
= \frac{\delta c_g^2}{c_g^2} + \delta \beta (1 + \frac{\delta P_{\rm gas}}{P_{\rm gas}}) 
\label{eqn:ct_cg},
\end{eqnarray}
\noindent
which can also be expressed in the following form:
\begin{eqnarray}
\frac{\delta c_T^2}{c_T^2}
&\approx& \frac{\delta \Gamma_1}{\Gamma_1} + \frac{\delta T}{T} - \frac{\delta \mu}{\mu} + \delta \beta (1 + \frac{\delta P_{\rm gas}}{P_{\rm gas}})
\label{eqn:ct_beta}
\end{eqnarray}
\noindent
by using the relation $P/\rho \propto T/\mu$.
Although $P/\rho \propto T/\mu$ is the same form as
the equation of state for an ideal gas, we are assuming that
all non-ideal effects are incorporated in the term $\mu$.
While $\mu$ is therefore not strictly the mean molecular weight
in the above equations,
this allows us to express the sound speed as above
and allows
Equations~(\ref{eqn:ct_cg}) and (\ref{eqn:ct_beta}) 
to be valid in non-ideal gas situations.
It is in fact a standard procedure to modify $\mu$
in the region of our study 
because the main non-ideal effect in this region is ionization.

As can be seen from Equations~(\ref{eqn:ct_cg}) and (\ref{eqn:ct_beta}), 
$\delta c_T^2/c_T^2$ results from not only the magnetic effects
but also the difference in the sound speed,
which represents the difference in the gas structure.
When the magnitude of $\delta \beta$ is comparable to
the magnitudes of other differences in Equation~(\ref{eqn:ct_beta}),
$\delta c_T^2/c_T^2 - \delta \Gamma_1/\Gamma_1$ 
should not be interpreted as the temperature difference.
Nevertheless,
if $\delta \beta$ can be determined and 
if $\delta P_{\rm gas}/P_{\rm gas} \ll 1$,
we can obtain the relative difference in both sound speed and stratification:
\begin{eqnarray}
\frac{\delta c_g^2}{c_g^2} 
&\approx& 
\frac{\delta c_T^2}{c_T^2} - \delta \beta 
\label{eqn:cg2} \\
\frac{\delta T}{T} - \frac{\delta \mu}{\mu}
&\approx&
\frac{\delta c_T^2}{c_T^2} - \frac{\delta \Gamma_1}{\Gamma_1} - \delta \beta
\label{eqn:T_mu}
\end{eqnarray}

\subsection{Procedure to infer Magnetic and Thermal Structures}
\label{subsec:calib_scheme}

Investigations of the  structure of active regions are usually based on
the assumption that
the ``sound-speed'' inversion results are a 
good measure of the actual sound-speed difference between 
active and quiet regions.
Under such assumption,
$\delta c^2/c^2$ is considered as due entirely to the magnetic fields
while $\delta c^2/c^2 - \delta \Gamma_1/\Gamma_1$ is interpreted
as difference in temperatures.
Although such interpretations may give qualitatively correct results,
they are not quantitatively correct,
as demonstrated by Equations~(\ref{eqn:ct_cg}) and (\ref{eqn:ct_beta}).
While it
is true that in general a lower wave speed implies a lower temperature, the
quantitative relationship is less direct. 
Since $\delta\Gamma_1/\Gamma_1$ follows
the distribution of the magnetic field (see Figure~\ref{fig:1}), 
we explore the possibility of
obtaining a relation between $\delta\beta$
and  
$\delta\Gamma_1/\Gamma_1$.
In the region where ring-diagram inversion is valid
(approximately $0.975R_\odot$ to $0.992R_\odot$),
the effort is complicated by the fact that the region is an ionization zone
and that $\Gamma_1$ changes as a result of ionization.
The presence of magnetic fields will change the temperature profile,
which in turn leads to shifting of ionization zones and
thereby a change in $\mu$ at each depth.
Consequently,
any relation we find between $\delta\beta$ and $\delta\Gamma_1/\Gamma_1$ 
is likely to depend on depth.

For the purpose of determining the relation between $\delta\beta$ and
$\delta\Gamma_1/\Gamma_1$,
we have constructed a large number of magnetic models
by adjusting the three parameters ($B_0$, $M_{Dc}$, and $\sigma$) 
of the magnetic profile,
as described in Section~\ref{sec:mdl}.
The values of these parameters range from
10 to 1000G for $B_0$,
$0.9R_\odot$ to $0.998R_\odot$ for the location of peak
and 0.1 to 0.9 for the width parameter.
In order to validate any relation that we might obtain,
we have also constructed several models with more complex
magnetic fields using multiple Gaussian profiles.

\begin{figure}
\centerline{
\includegraphics[width=0.64\linewidth,angle=90]{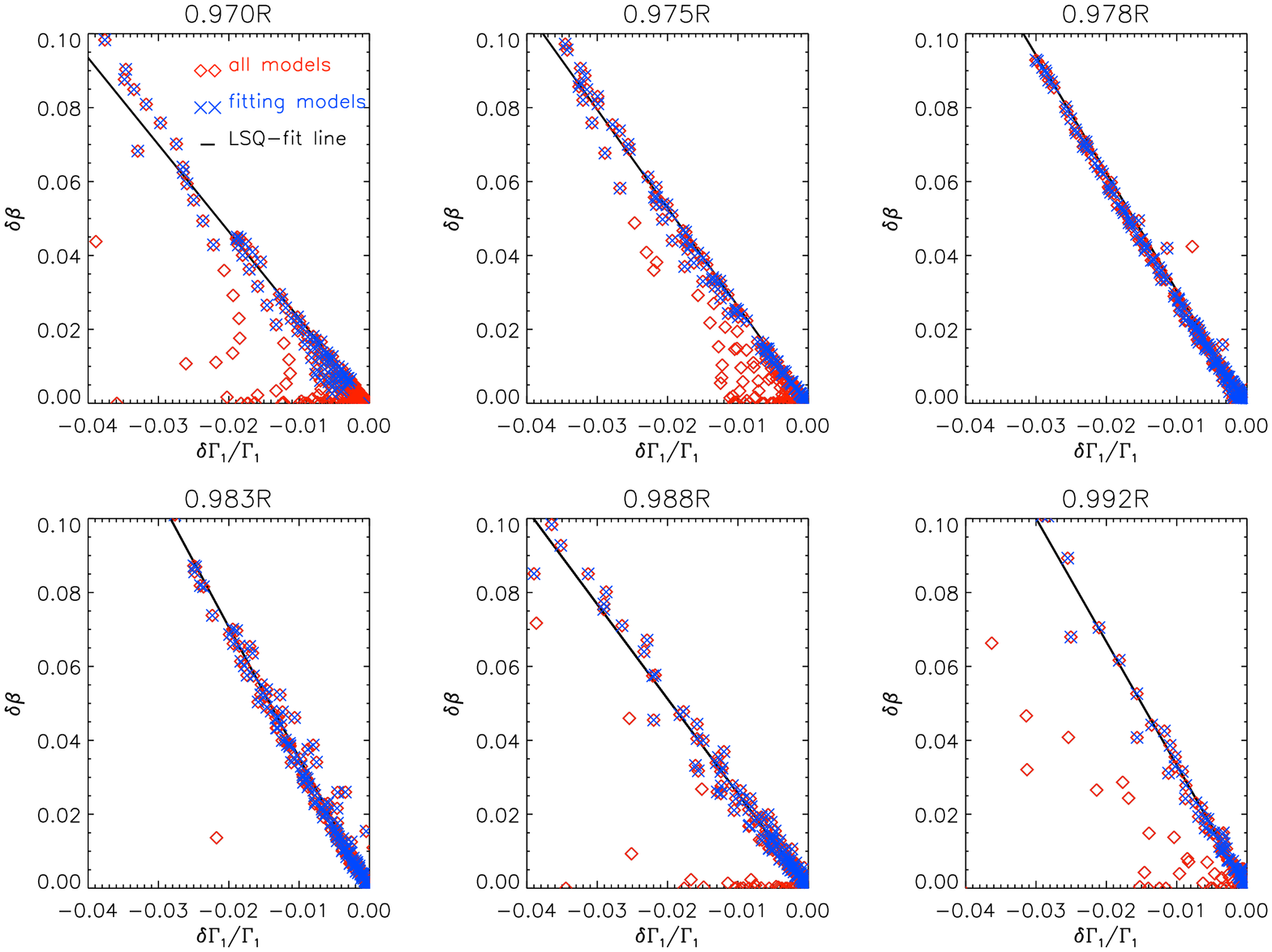}}
\centerline{
\includegraphics[width=0.64\linewidth,angle=90]{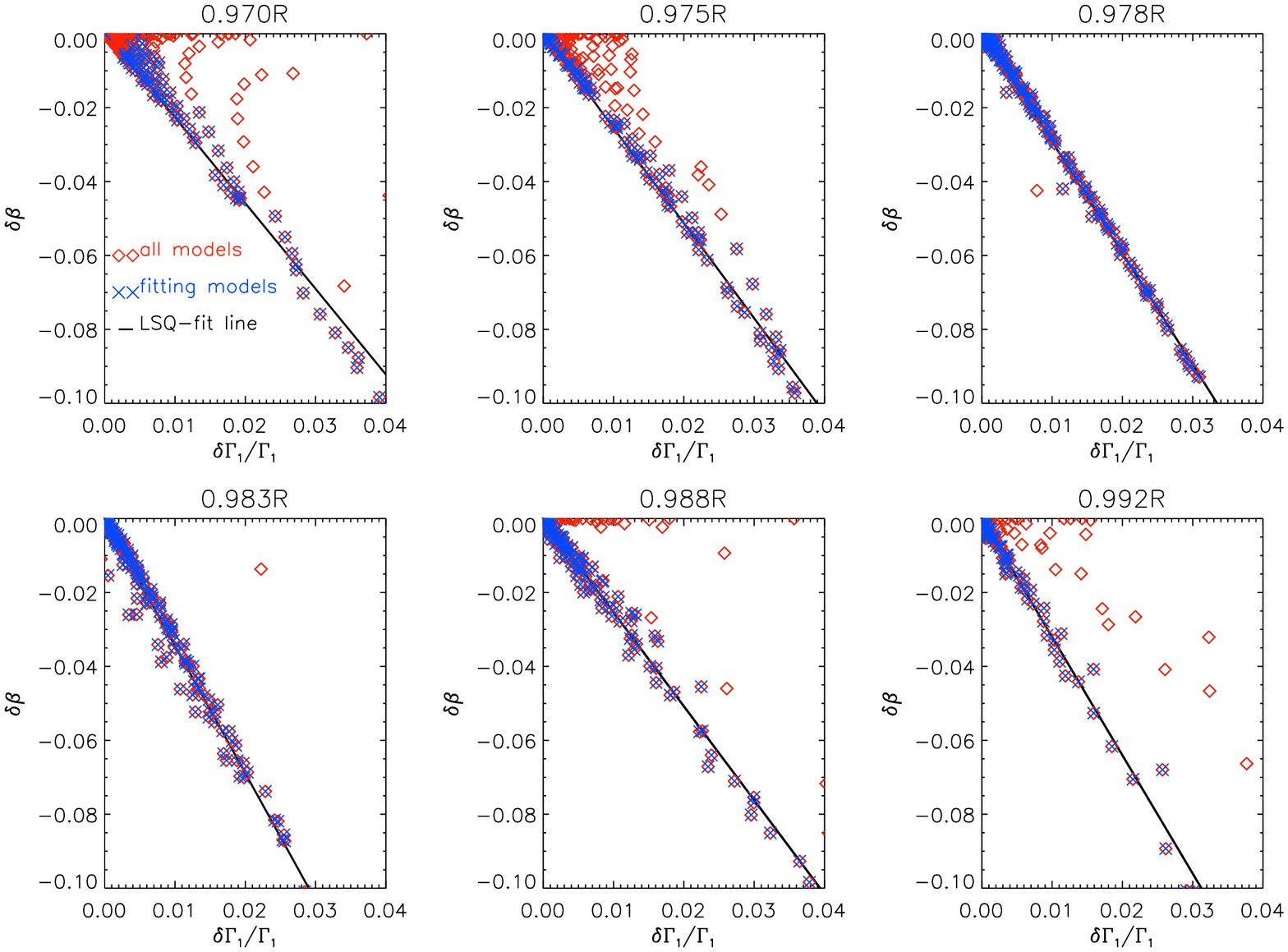}}
\caption{The  $\delta \Gamma_1/\Gamma_1$ versus $\delta \beta$ relation
at different depths.
The upper two rows are the relations of magnetic model $-$ non-magnetic model,
and the lower two rows are the relations of non-magnetic model $-$ magnetic models.
Each diamond represents one model pair,
or equivalently, a unique magnetic profile.
The line is the result of a  least squares fit (LSQ-fit) 
obtained after excluding
some stray points.
The crosses are the ones included for the fit.}
\label{fig:fit}
\end{figure}


Figure~\ref{fig:fit} shows $\delta \Gamma_1/\Gamma_1$ versus $\delta \beta$ 
at different depths.
Since the values of $\delta \Gamma_1/\Gamma_1$ obtained from 
the inversions 
by Basu, Antia, and Bogart (2004)
range from negative to positive, 
we need to investigate the relation between 
$\delta\beta$ and $\delta\Gamma_1/\Gamma_1$
for both positive and negative $\delta\Gamma_1/\Gamma_1$.
For this purpose,
we created negative (positive) $\delta\Gamma_1/\Gamma_1$ by
using non-magnetic (magnetic) models as the reference 
and magnetic (non-magnetic) models as the test model.
In Figure~\ref{fig:fit},
we plotted negative (positive) $\delta\Gamma_1/\Gamma_1$ versus $\delta\beta$
in the upper (lower) two rows.
In all panels, each diamond symbol represents the computed values 
($\delta \Gamma_1/\Gamma_1$, $\delta \beta$) of a model pair at the depth
indicated above each panel.
The scale of abscissa
is chosen to be comparable to 
the values of $\delta \Gamma_1/\Gamma_1$ from the inversions.
The figure shows that while there is a tight relationship between
$\delta\beta$ and $\delta \Gamma_1/\Gamma_1$
in the region $0.975R_\odot < r < 0.990R_\odot$,
where the inversion is reliable,
there is often no
unique relation, but only an envelope, 
in the deeper and shallower layers.
A closer examination of the models associated with the points 
located away from the envelope reveals that 
the $\delta \Gamma_1/\Gamma_1$ and
$\delta c_T^2/c_T^2 - \delta \Gamma_1/\Gamma_1$ profiles
resulting from these models
are very different from the inversion results.
For instance,
the point in the panel of $0.97R_\odot$
whose $|\delta \Gamma_1/\Gamma_1|$ has a magnitude of approximately 0.012 
and $|\delta \beta|$ roughly 0.005
belongs to the $\delta \Gamma_1/\Gamma_1$ profile
with a peak magnitude greater than 0.1 at $0.985R_\odot$ 
({\it cf.} Figure~\ref{fig:est_mdl}(g)).
Other points that locate even further away from the envelope are produced by 
those $\delta \Gamma_1/\Gamma_1$ profiles with even larger peak magnitudes.
However, 
the magnitudes of the inverted $\delta \Gamma_1/\Gamma_1$ shown in
Basu, Antia, and Bogart (2004)
are all less than 0.04.
Thus, 
these stray points appear to result from unrealistic magnetic configurations,
and hence, we do not use them to determine the relation.
After removing these stray points,
we applied a least squares fit (LSQ fit) to the rest of data to obtain 
a linear relation at each selected depth.  
The straight line in each panel is our fitted line, and
is the relation between $\delta\beta$ and
$\delta\Gamma_1/\Gamma_1$ at that depth to be used to
infer $\delta \beta$ from the solar inversion results.

\subsection{Validating the $\delta\beta$ -- $\delta\Gamma_1/\Gamma_1$ Relation}
\label{subsec:calib_valid}
%
%
%
Using the known $\delta \Gamma_1/\Gamma_1$
between the test models and the reference models, 
we first determined $\delta\beta$ between the models 
based on 
the $\delta\beta$ -- $\delta\Gamma_1/\Gamma_1$ relation.
Once $\delta \beta$ is determined,
we then ventured to examine whether, and how accurately,
$\delta c_g^2/c_g^2$ and $(\delta T/T-\delta \mu/\mu)$
can be inferred from 
$\delta c^2_T/c_T^2$, $\delta \Gamma_1/\Gamma_1$ and $\delta \beta$,
as explained in Section~\ref{subsec:result1}
[{\it cf.} Equations~(\ref{eqn:cg2}) and (\ref{eqn:T_mu})].
These quantities can then 
be compared with the exact differences between the models
to assess the errors of the estimations.
While $\delta \mu/\mu$ is negligible in the deeper regions where gas
is almost fully ionized,
it has to be taken into account at the vicinity of ionization zones,
which are within the region we are probing. 
$\delta T/T$ and $\delta \mu/\mu$ can  be separated only if it is assumed
that we know the equation of state of solar matter correctly.
We therefore, chose to determine $(\delta T/T - \delta \mu/\mu)$
rather than $\delta T/T$ in this
paper to avoid possible errors from the equation of state.

Since the relation is obtained by 
removing several unrealistic models and
using fairly simple magnetic-field configurations,
we also need to assess the accuracy of the relationship under 
these excluded magnetic conditions.
Hence,
we applied the aforementioned steps to 
the test models with multiple-Gaussian magnetic profiles and
the test models that result in those stray points in Figure~\ref{fig:fit}.
We also applied the procedure to infer $\delta c_g^2/c_g^2$,
$\delta\beta$ and $(\delta T/T - \delta \mu/\mu)$
between two magnetic models, 
that is, both test and reference models are magnetic.

\begin{figure}
\centerline{
\includegraphics[height=1.2\linewidth]{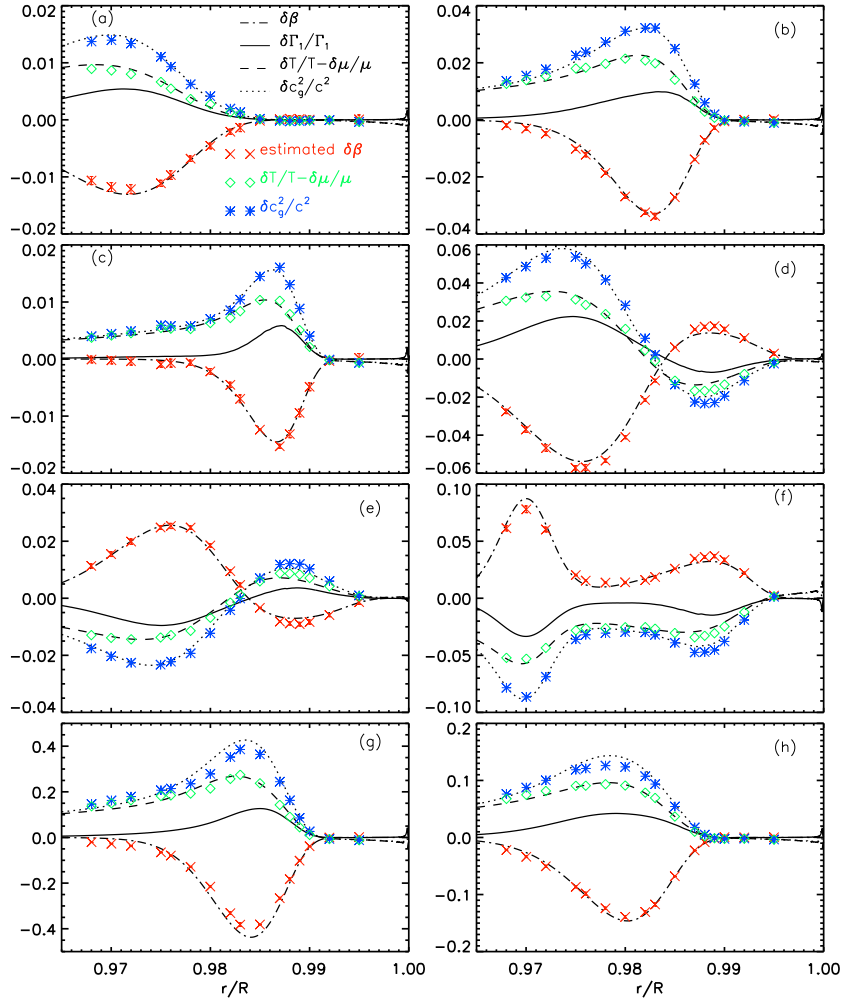}}
\caption{ The result of applying the $\delta\Gamma_1/\Gamma_1$ -- $\delta\beta$
relation to models.
The symbols are
the $\delta\beta$,  $(\delta T/T-\delta \mu/\mu)$, and $\delta c_g^2/c^2$
determined by using the known values of
$\delta\Gamma_1/\Gamma_1$, $\delta c^2_T/c^2_T-\delta\Gamma_1/\Gamma_1$,
and Equations~(\ref{eqn:cg2}) and (\ref{eqn:T_mu}).
The exact values of the determined quantities
are shown as continuous lines.
}
\label{fig:est_mdl}
\end{figure}


The results of the exercise are shown in Figure~\ref{fig:est_mdl}.
Panels (a), (b), (c) are the results obtained by applying
our relation to models with simple  magnetic fields concentrated at 
different depths;
panels (d) and (e) show two examples when 
both test and reference models are magnetic;
(f) is a case of complex magnetic fields which 
is created by a double-Gaussian profile;
and panels (g) and (h) show two model pairs that are excluded from our fitting.
We can see that the reconstructed $\delta \beta$ (the red crosses) closely match
the computed values (the dot-dash line).
Even in the cases where $\delta\beta$ is reasonably complex,
the simple relation still re-creates the $\delta \beta$ to an accuracy that
is sufficient to apply to the solar results.
As can be seen in the same figure,
our estimated $\delta c_g^2/c_g^2$ (blue stars)
and $(\delta T/T - \delta \mu/\mu)$ (green diamonds),
still follow the exact values, but
show slightly larger deviations in several cases, 
which is an indication that
$\delta P_{\rm gas}/P_{\rm gas}$ between the test and reference models
is not negligible in these cases.
The results in panel (g) demonstrate
how much the estimated and computed $\delta \beta$ would differ
in the excluded model pairs.

\section{Application to Solar Data}\label{sec:sun}

\begin{table*}
\caption[]{Properties of the different pairs of regions
that were analyzed}
\label{tab:mai}
\begin{tabular}{lcccccc}
\hline
{Pair}&
Lat.&
CM$^a$ Lon.&
Mag. Index&
NOAA& 
Type& 
Max. area \\
{No.} & 
\multicolumn{1}{c}{(deg.)} & 
\multicolumn{1}{c}{(deg.)} &
\multicolumn{1}{c}{(Gauss)} &
\multicolumn{1}{c}{\ } & 
\multicolumn{1}{c}{(at CMP$^b$) }&
\multicolumn{1}{c}{(millionths)} \\
\hline
\phantom{0}1 & 7N & 016 & 19.9 & 8040 & $\beta$  & \phantom{0}150 \\
 &      & 341 & 0.2 \\
\phantom{0}2 & 16S & 222 & 23.3 & 9904 & $\beta$ & \phantom{00}60 \\
 & & 242 & 2.7 \\ 
\phantom{0}3  & 11S & 195 & 26.8 & 9896 & $\alpha$ & \phantom{0}110 \\
 & & 205 & 2.0 \\
\phantom{0}4  & 14S & 105 & 53.2 & 8518 & $\beta$ & \phantom{0}170 \\
 &       & 075 & 0.9 \\
\phantom{0}5  & 18N & 180 & 56.3 & 9899 & $\beta$ & \phantom{0}220 \\
 & & 240 & 2.2  \\
\phantom{0}6  & 21S & \phantom{0}82 & 68.4 & 8193 & $\beta$ & \phantom{0}290 \\
 &       & 067 & 0.6 \\
\phantom{0}7  & 19N & 215 & 81.5 & 9893 & $\beta\gamma\delta$ & \phantom{0}490 \\
 & & 255 & 0.7 \\
\phantom{0}8 & \phantom{0}4N  & 013 & 86.9 & 9914 & $\beta$ & \phantom{0}260 \\
 & & 028 & 1.1  \\
\phantom{0}9  & 20N & 204 & 108.5 & 9901 & $\beta\gamma$ & \phantom{0}350 \\
 & & 249 & 0.7  \\
10 & 15S & 150 & 125.8 & 9906 & $\beta\gamma\delta$ & \phantom{0}850 \\
 & & 120 & 2.8 \\ 
11 & 20N & 071 & 146.6 & 9026 & $\beta\gamma\delta$ & \phantom{0}820 \\
 &       & 126 & 0.9 \\ 
12 & 19N & 147 & 241.6 & 9393 & $\beta\gamma\delta$ & 2440 \\
 & & 207 & 1.2 \\ 
\hline
\end{tabular}

$^a$ Central Meridian\\
$^b$ Central Meridian Passage

\end{table*}


Once the reliability of the relation was established, we applied it 
to the inversion results.
Although the estimated $\delta T/T - \delta \mu/\mu$ and $\delta c_g^2/c_g^2$ 
are qualitatively correct in our model tests,
the larger uncertainties in the real solar data may result in
even more erroneous estimations.
Hence, 
we shall only focus on $\delta \beta$ in the application to the solar data.

Basu, Antia, and Bogart (2004)
determined
$\delta c_T^2/c_T^2$ and $\delta \Gamma_1/\Gamma_1$
of twelve selected active region(AR) -- quiet region(QS) pairs
by using an inversion procedure with the frequencies determined by
a ring-diagram analysis.
The selected active regions were tracked while
they were crossing the central meridian (CM).  
The QS in each pair was selected such that
it also crossed the central meridian at the same latitude as the AR
within the same Carrington rotation.  
In other words, the QS either preceded or
followed the AR. 

The  magnetic-field strengths of the ARs scaled a wide range.  
The magnetic field strength of each region is
defined by a magnetic activity index (MAI), which is an average
of the absolute values of the strong fields ($|B_z| \ge 50$ G).  
Details of how the MAI for
a region is calculated can be found in  Basu, Antia, and Bogart (2004).
Some of the properties of the regions
are described in Table~\ref{tab:mai}.
The table is arranged in order of increasing MAI of the
active region. 

\begin{figure}
\centerline{
\includegraphics[height=1.2\linewidth]{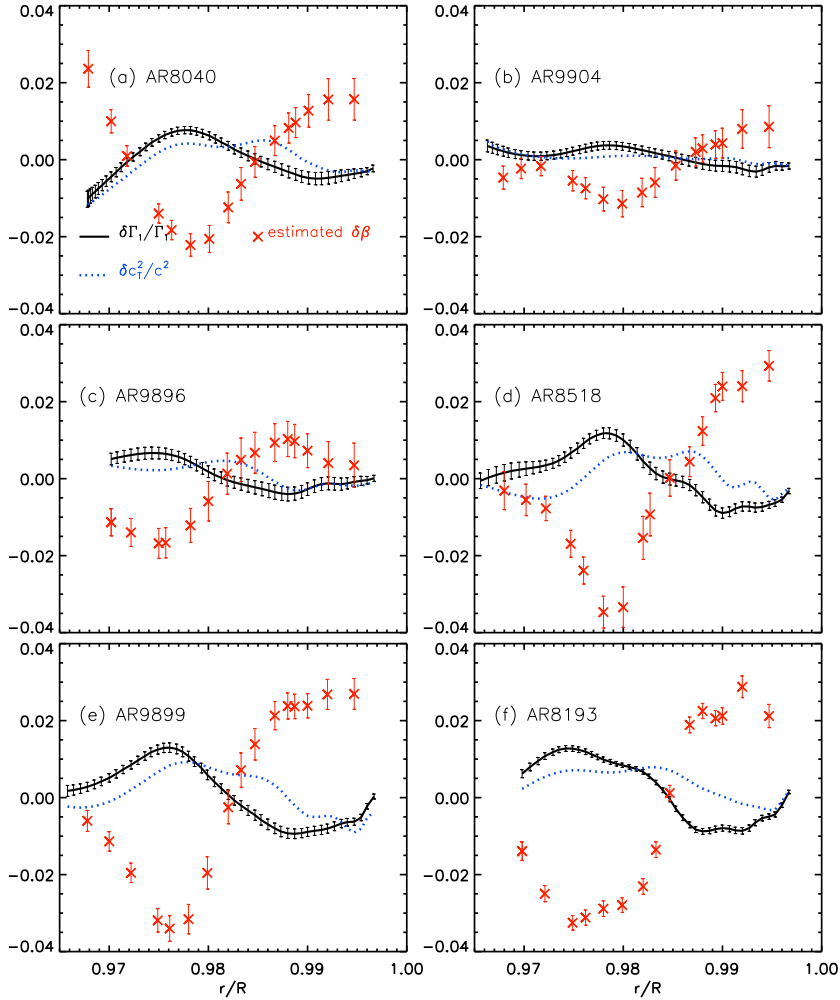}}
\caption{
The estimated $\delta \beta$
between different pairs of
solar active and quiet regions
are plotted as red crosses along with the error bars.
Also shown are the primary
inversion results for $\delta\Gamma_1/\Gamma_1$ (solid line) and
$\delta c^2_T/c^2_T$ (dotted line) 
obtained by Basu, Antia, and Bogart  (2004).
Because the error bars of $\delta\Gamma_1/\Gamma_1$ and
of $\delta c^2_T/c^2_T$ are of similar magnitude,
only the former are plotted for the sake of clarity.
}
\label{fig:solar1}
\end{figure}

\begin{figure}
\centerline{
\includegraphics[height=1.2\linewidth]{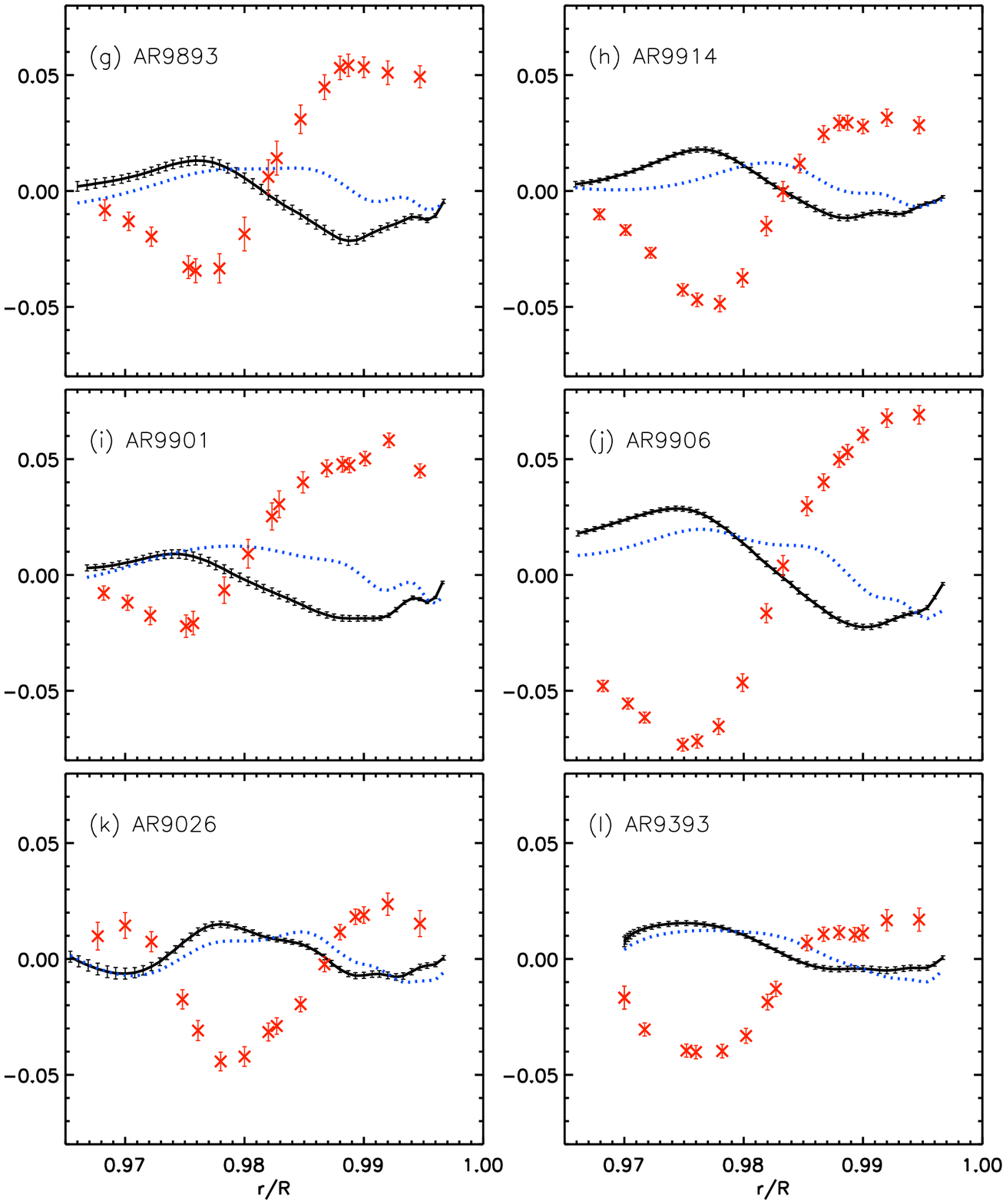}}
\caption{
The estimated $\delta \beta$
between different pairs of
solar active and quiet regions
are plotted as red crosses along with the error bars.
Also shown are the primary
inversion results for $\delta\Gamma_1/\Gamma_1$ (solid line) and
$\delta c^2_T/c^2_T$ (dotted line) 
obtained by Basu, Antia, and Bogart  (2004).
Because the error bars of $\delta\Gamma_1/\Gamma_1$ and
of $\delta c^2_T/c^2_T$ are of similar magnitude,
only the former are plotted for the sake of clarity.
}
\label{fig:solar2}
\end{figure}


The estimated values of $\delta\beta$ (red crosses),
along with the inversion results and error bars,
for the different AR -- QS pairs are plotted in 
Figures~\ref{fig:solar1} and \ref{fig:solar2}.  
The error bars are computed by propagating
the errors from fitting and from the inversion results of 
$\delta\Gamma_1/\Gamma_1$
through the  $\delta \beta$ - $\delta\Gamma_1/\Gamma_1$ relationship
derived in Section~\ref{subsec:calib_scheme}.
The figures show that in the region above approximately $0.985R_\odot$,
$\delta \beta$ generally increases with increasing MAI.
The only notable exceptions are AR$\,9026$ and AR$\,9393$.
The anomalously low near-surface $\delta \beta$ of the strong active regions 
AR$\,9026$ and AR$\,9393$ is worth commenting upon. 
These two regions have the largest number of
the flares among all the regions studied, 422 for AR$\,9026$ and 
568 for AR$\,9393$.
The next highest flare index is only 129 for AR$\,9906$. 
It is therefore quite likely
that the flaring activity and the small $\delta\beta$ are related. However, we
need to study a much larger sample before we can draw any definite conclusions
about the relation between flaring activity and $\delta \beta$
in the shallow subsurface layer.
Another noticeable feature is that
in the region below approximately $0.985R_\odot$,
$\delta\beta$ becomes negative.
The magnitude of this negative $\delta\beta$ in the deeper region
is often comparable to
that of the positive $\delta\beta$ above $0.985R_\odot$.

Hence, the results indicate
that 
the strongest magnetic effects,
which we define as the largest change in $\beta$,
in ARs are generally located
in a shallow region above approximately $0.985R_{\odot}$
and that 
the effect of magnetic fields ({\it i.e.}, $\beta$) in ARs become smaller 
than $\beta$ in QSs
in the deeper layers.
In addition,
the roughly comparable magnitudes of 
near-surface and deeper-layer $|\delta \beta|$
seem to imply that
the stronger the AR is at the surface
the smaller its $\beta$ is in the deeper layers.
However,
it could also be that
the QSs chosen to pair with the weaker ARs have larger $\beta$ than
the QSs chosen for the stronger ARs.

Because $\beta$ is defined as the ratio between magnetic pressure and
gas pressure ($P_{\rm mag}/P_{\rm gas}$),
it is an indicator of the competition between the magnetic and gas effects.
Hence, the magnitude of $\beta$ simply indicates 
whether the gas or magnetic effects dominate in a region 
but not the actual magnitude of the magnetic fields.
In other words,
a smaller $\beta$ could result from either a genuine weaker magnetic pressure
or a greater gas pressure.

\section{Interpretation of Inferred Results}\label{sec:interpretation}

To verify the possibilities proposed in the previous section,
we 
examined $\delta \beta$
between regions that have different MAIs and/or
are observed at different times and/or locations.
The main goal is to probe
the distribution profiles of $\beta$ beneath an AR and a QS.
Specifically, we aim to 
reveal whether there is indeed non-negligible $\beta$ in the QSs
and how $\beta$ of both ARs and QSs change over the depth.
We also wish to 
investigate how the profile of $\beta$ varies
with time, location,
and photospheric magnetic-field strength of the region observed.
In addition,
while we need to study more cases
to confirm whether ARs with prolific flares
are often linked to small surface $\beta$,
we wish to propose and discuss possible 
connections between the flaring activities
and the subsurface $\beta$.

The regions paired for this investigation are:
AR$\,9393$ (2001 March, MAI $=$ 241.6G) versus 
AR$\,9026$ (2000 June, MAI $=$ 146.6G),
AR$\,9906$ (2002 April, MAI $=$ 125.8G) versus 
AR$\,8518$ (1999 April, MAI $=$ 53.2G),
AR$\,9914$ (2002 April, MAI $=$ 86.9G) versus 
AR$\,8040$ (1997 May, MAI $=$ 19.9G), and
AR$\,9914$ versus AR$\,9904$ (2002 April, MAI = 23.3G).
For each pair,
we computed the differences between the active regions ({\it i.e.}, AR1--AR0)
and between their companion quiet regions ({\it i.e.}, QS1--QS0).
We also computed the difference of AR1--QS0 
so that the consistency  of the results can be checked.

\begin{figure}
\centerline{
\includegraphics[height=1.2\linewidth]{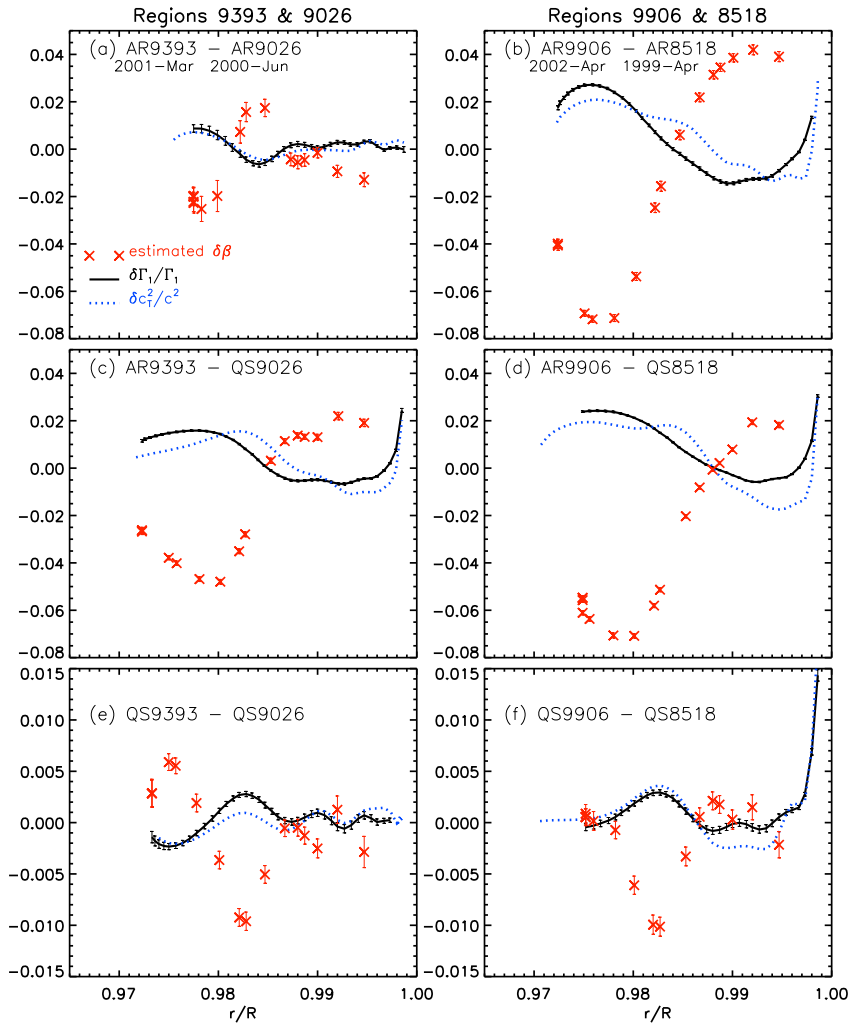}}
\caption{
The inversion results and the estimated $\delta \beta$
between different regions,
as denoted in the plots.
The date of observation is indicated under each corresponding region.
The first row displays the results between two different active regions (ARs),
the second row is between one of the ARs 
and the quiet-Sun region (QS) of the other one, and
the bottom row shows the QS of the two active regions.
}
\label{fig:solar_difAR1}
\end{figure}

\begin{figure}
\centerline{
\includegraphics[height=1.2\linewidth]{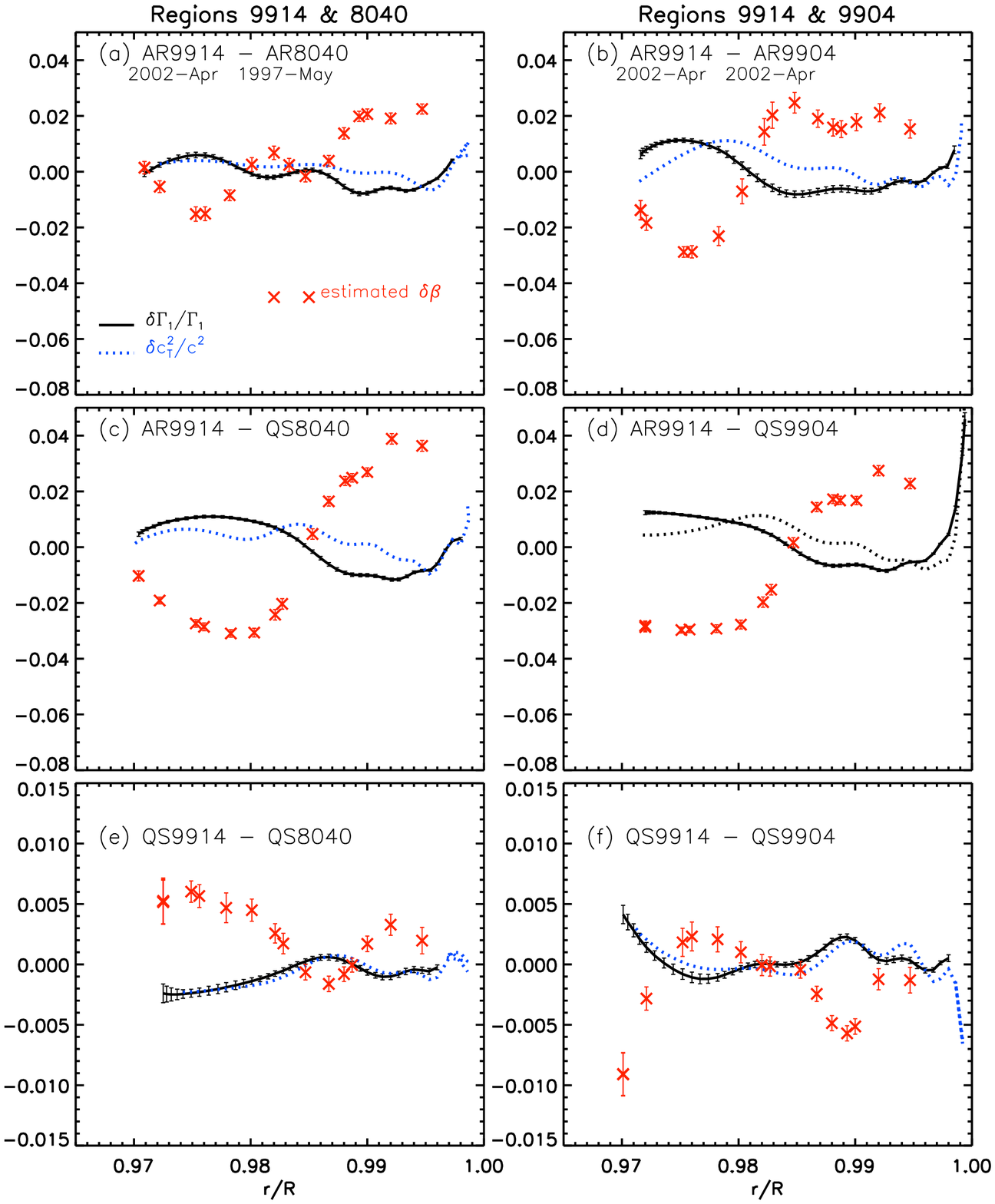}}
\caption{
The inversion results and the estimated $\delta \beta$
between different regions,
as denoted in the plots.
The date of observation is indicated under each corresponding region.
The first row displays the results between two different active regions (ARs),
the second row is between one of the ARs
and the quiet-Sun region (QS) of the other one, and
the bottom row shows the QS of the two active regions.
}
\label{fig:solar_difAR2}
\end{figure}

$\delta \beta$ of AR1--AR0 is to investigate 
whether or not the distribution of $\beta$
differs between different ARs.
The examination of
$\delta \beta$ between two QSs is firstly to verify
the non-negligible magnitude of $\beta$ suggested by the results
in Figures~\ref{fig:solar1} and \ref{fig:solar2} and secondly to determine 
whether the magnetic fields in QSs are uniform or different
for the different quiet regions.
The results are shown in 
Figures~\ref{fig:solar_difAR1} and \ref{fig:solar_difAR2}.
The top and bottom rows illustrate the differences between 
different ARs and between different QSs, respectively.
The results of AR1--QS0 for the consistency check are plotted in the middle row.
%

\subsection{Magnetic Effects beneath Active Regions}
\label{subsection:interp_AR}
We can see from the top row of Figures~\ref{fig:solar_difAR1} and
\ref{fig:solar_difAR2} that
in the region below approximately $0.985R_\odot$,
the $\beta$ of ARs with larger MAI is indeed smaller than 
that of ARs with smaller MAI. 
The comparison of different QSs as illustrated in the bottom row
shows that $\beta$ in these QSs does not have explicit correlation
with MAI of their companion ARs.
For instance,
while QS$\,9906$ is weaker than QS$\,8518$,
QS$\,9914$ is clearly stronger than QS$\,8040$.

A tentative picture of the magnetic structure beneath an AR that
we can conjure up from these results is
that the magnetic effects of the ARs are most prominent
in a shallow region above $0.985R_\odot$.
In the deeper layers, 
the effects drop off more significantly in a strong AR than in a weak one.
The reason could be that the magnetic fields near the surface
reduce the convective motion and thus 
the gas pressure and density in the region, 
which can lead to an increased concentration of gas density and 
increased gas pressure
below the magnetized area.
The stronger the magnetic fields are, 
the higher the increased concentration of gas in the deeper layers would be.
Such increased $P_{\rm gas}$, therefore, could result in lower $\beta$.

\subsection{Magnetic Effects beneath Quiet Regions}
\label{subsec:interp_QS}
The first common feature we can see from the bottom row of
Figures~\ref{fig:solar_difAR1} and \ref{fig:solar_difAR2} is that
$|\delta \beta|$ of QSs is smaller than that of ARs.
In principle,
there are two possibilities for the smaller $|\delta \beta|$.
It could be either the {\em variation} or the {\em magnitude} of $\beta$
is smaller in QSs than in ARs.
We can rule out the second one because of the following reason. 
The magnitudes of $\delta \beta$ in the results of AR1--AR0
provide a lower limit for $\beta$ in the active regions
({\it i.e.}, ${\rm max}(\beta_1, \beta_0) \ge |\beta_1 - \beta_0|$).
The large negative $\delta \beta$ seen in all AR--QS plots in
Figures~\ref{fig:solar1} to \ref{fig:solar_difAR2}
indicate that the deep-layer $\beta$ in the quiet regions should not be 
smaller than that in any of the active regions.
As an example,
AR$\,9906$ -- AR$\,8518$ and AR$\,8518$ -- QS$\,8518$ imply that 
$\beta_{\rm AR}\,8518 \ge 0.075$ and, consequently,
$\beta_{\rm QS}\,8581 \ge 0.075+0.04$.
Therefore, we can deduce from the plots that
there are indeed non-negligible $\beta$ in the quiet regions
and that
the profile of $\beta$ varies from region to region
albeit being more uniform and stable than that in the active regions.
It should be noted that the QSs were selected to precede or follow the selected
ARs, and hence, this variation between QSs is probably an indication that
the evolution of magnetic fields in an AR differs from AR to AR. This is
confirmed by the results seen for the pair QS$\,9914$ -- QS$\,9904$, which are
QSs associated with two different active regions during the same Carrington
rotation. Because of the way the QSs were selected, such studies could be used
to study the evolution of active regions.

The conclusions we can deduce from the examination of QSs
are that
there are non-uniform and significant magnetic structures in quiet regions.
Although the results seem to suggest that the temporal variation is
greater in the deeper layers than in the shallower layers,
the number of our test group is too small to draw a definite conclusion.

\subsection{The Anomaly in AR$\,9026$ and AR$\,9393$}
\label{subsec:interp_flare}
Next, we would like to address the small inferred $\delta \beta$ in the two ARs 
(AR$\,9026$ and AR$\,9323$)
that have the highest MAI and flare number among our studied regions.
As the number of such cases in our current data set is too small for us 
to deduce statistically valid statements,
we shall only propose in the following several possible explanations 
based on relevant theories and
recent observational results:
\smallskip

\noindent{\bf {\it i)} $P_{\rm gas}$ is higher in the ARs with 
higher flaring activities, which leads to smaller $\beta$:}

Observations have shown that
the magnetic fields at the photosphere level remain roughly unchanged
before and after flaring events,
which means that the magnetic pressure ($P_{\rm mag} \equiv B^2/8\pi$)
should not be affected much by the flares.
Hence,
the small $\beta$ is more likely due to an increase in 
gas pressure and density in the shallow layers.
One possible explanation for the increased gas density
is that
the amount of flux coming up from the convection zone
may be larger
in the flaring active regions than in the non-flaring ones.
Based on the storage model,
which is a commonly accepted solar eruption model,
the emerging flux accompanied with foot-point movements of coronal field lines
can cause magnetic stress to build up in the corona and eventually lead
to eruption 
(Priest and Forbes 2000; Lin, Li, and Basu 2003).
Another possible source of the increased gas density is 
from the material flowing down from the corona after the flares.
However, 
although it has indeed been reported by helioseismology that
some large flares can cause detectable ripples at the photospheric level
(Kosovichev and Zharkova 1995, 1998; Zharkova and Kosovichev 1998),
the mass flow from the corona,
where the plasma density is approximately $10^8$ times smaller than 
that at the photosphere,
is usually insufficient to affect surface $P_{\rm gas}$ (hence $\beta$)
by a factor of two or three, as suggested by our results
({\it cf.} Figure~\ref{fig:solar2}).
In addition, observations have found
that although
up-flows and down-flows happen consecutively at the chromospheric level
right after a flare,
the temperature and gas density eventually return to the quiet-Sun values
({\it e.g.}, Raftery et al. (2008)).
Based on the conservation of mass,
we may expect similar situation also occur at the photosphere level.
\smallskip

\noindent{\bf {\it ii)} The relation between $\delta \beta$ and 
$\delta \Gamma_1/\Gamma_1$ derived from our models may not be suitable
for the shallow layers of such highly dynamic and eruptive ARs:} 

In our models, the only source to affect $\Gamma_1$ is 
the magnetic field. 
However, in an explosive active region,
where many dynamic activities take place,
the variations in $\Gamma_1$ can come from many sources.
Specifically,
during a flaring event, 
the material in the low atmosphere is heated by the
energetic particles coming down from the corona. 
While the total mass of the down-flow material is small 
relative to the photospheric density,
the heating and the interaction with the energetic particles
can change the state of ionization, entropy,
and even the equation of state itself
at the surface layer.
All of these effects can change $\Gamma_1$,
and thus become part of the source for 
the $\delta \Gamma_1/\Gamma_1$ in our solar data.
As a result, 
$\delta \beta$ inferred from the relation derived from the models
without these effects
may not reflect the actual value.
For instance,
the heating may counter-act the cooling effects from the magnetic fields
and cause a smaller magnitude of $\delta \Gamma_1/\Gamma_1$,
which leads to a smaller inferred $\delta \beta$ at the surface.
In contrast,
in the deeper region
where the effect of the energetic particles is less and
the thermal structure is better described by the structure of our models,
the inferred $\delta \beta$ would be closer to the actual value,
which could be the reason why 
the deeper-layer $\delta \beta$ of AR$\,9026$ and AR$\,9393$
is a factor of two to four larger than the surface-layer values.
\smallskip

The above hypotheses may be verified by studying
a number of active-region pairs
of which both have similar MAIs but
one has many flares and the other has zero flares.
This exercise would first confirm whether 
the small $\beta$ near the surface is indeed
a general trend in the flaring ARs.
If it is positively confirmed,
a large $|\delta \beta|$ in the deep layers would be an indication that
there is likely to be emerging flux in the flaring ARs to cause the
small $\beta$.
In contrary, if we find that $|\delta \beta|$ is small in the deep layers,
it implies that there is no significant difference in the magnetic structure
between flaring and non-flaring ARs, and, therefore, 
our second hypothesis may be the explanation.

\section{Modeling Subsurface Magnetic Structure}
\label{sec:model}

\begin{figure}
\centerline{
\includegraphics[width=0.9\linewidth]{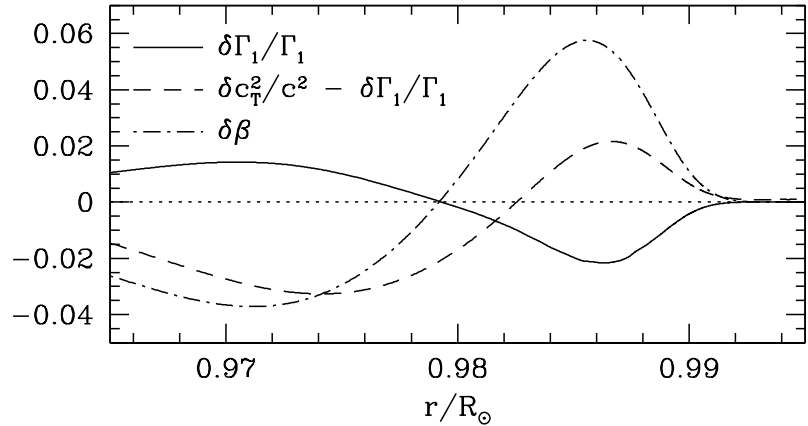}}
\caption{$\delta\Gamma_1/\Gamma_1$ and
$\delta c^2_T/c^2-\delta\Gamma_1/\Gamma_1$ between two models. 
The models were selected
such that the relative differences qualitatively mimic those
seen between solar active and quiet regions. Also shown is
$\delta\beta$ between the two models.
}
\label{fig:cmp}
\end{figure}

After confirming the reliability of our strategy,
we ventured to investigate if we may be able to tune
our simple one-dimensional magnetic models
to reproduce the main features commonly seen in the inversion data.
Specifically,
the two main features we aim to reproduce are
$\delta c_T^2/c_T^2 - \delta \Gamma_1/\Gamma_1$ being positive
above $\approx 0.985R_\odot$ but negative below that,
and an opposite profile for $\delta\Gamma_1/\Gamma_1$,
that is, negative close to the surface but positive in the deeper layers.
To produce positive $\delta c_T^2/c_T^2 - \delta \Gamma_1/\Gamma_1$
that resemble the inversion results,
our study indicates that
the magnetic fields must be both strong and concentrated. 
To create a large, positive $\delta \Gamma_1/\Gamma_1$
commonly seen between $0.97R_\odot$ and $0.985R_\odot$, 
the models suggest that
there has to be a negative $\delta \beta$ in that region. 
This can be
seen in Figure~\ref{fig:cmp}.
Although the shifting of ionization zones can also produce 
a positive $\delta \Gamma_1/\Gamma_1$,
such changes
are always localized around the ionization zone,
and the magnitudes of the changes are much smaller than 
what is observed.
Hence, our results imply that the quiet regions,
which have been considered as the non-magnetic references in the inversions,
could in fact contain non-negligible magnetic pressure in the deep layers.

\section{Summary}\label{sec:summary}

The aim of this investigation was to find a practical way to
determine the thermal and magnetic structure under active regions from
sound speed and $\Gamma_1$ results obtained by inverting frequency
differences between active and quiet regions.
We have shown that ``sound speed'' results obtained by inverting
frequency differences between active and quiet regions of the
Sun represent the ``wave speed'' ($c_T \equiv \sqrt{\Gamma_1 P_{T}/\rho}$)
and not the sound speed ($ c_g \equiv \sqrt{\Gamma_1 P_{\rm gas}/\rho}$) or
the Alfv\'{e}n speed.
Thus the results cannot be interpreted as being caused by differences
in either temperature or magnetic fields alone. 
A combination of the two is needed to interpret the results correctly.

Using solar models with magnetic fields, 
we have determined a simple relationship
between 
$\delta\Gamma_1/\Gamma_1$, 
which is the relative difference of adiabatic index,
and $\delta \beta$ ($\beta \equiv P_{\rm mag}/P_{\rm gas}$). 
$\delta\Gamma_1/\Gamma_1$ was chosen for this work because
its feature is well confined in the region of magnetic fields
and because it can be obtained unambiguously
from inversions.
We have had to
derive a separate relationship between the two quantities at each depth
because
the process of ionization, which changes $\Gamma_1$,
varies with depth.
The $\delta \beta$ - $\delta\Gamma_1/\Gamma_1$ relation can then
used to infer the subsurface $\delta \beta$.
The determined $\delta \beta$,
along with inversion results of
$\delta c_T^2/c_T^2$ and $\delta \Gamma_1/\Gamma_1$, 
can  be used to
determine $\delta c_g^2/c_g^2$ and $(\delta T/T-\delta\mu/\mu)$,
provided that
$\delta P_{\rm gas}/P_{\rm gas}$ between two structures 
is negligible ({\it i.e.}, $\delta P_{\rm gas}/P_{\rm gas} \ll 1.$). 
This method has been validated by using models.
However,
as the assumption $\delta P_{\rm gas}/P_{\rm gas} \ll 1.$ may not be suitable
in the Sun,
we only present in this paper $\delta\beta$ revealed by applying the method to 
the solar results obtained
by Basu, Antia, and Bogart (2004).

The application of $\delta \beta$ -- $\delta\Gamma_1/\Gamma_1$ relation to
the solar inversion results shows that, as expected,
different active regions behave quite differently. 
Nevertheless,
for all regions studied, $\delta\beta$ is always positive immediately
below the surface ( $ r > 0.985R_\odot$) but negative at deeper regions, 
which implies that
the inward decrease of $\beta$ is faster in an AR with larger MAI
than in one with smaller MAI.

The reason for the faster decrease of $\beta$ under
stronger ARs could be because the stronger magnetic
fields near the surface suppress convection more, resulting
in a larger build-up of gas ({\it i.e.}, higher density and gas pressure),
which leads to 
lower $\beta \equiv P_{\rm mag}/P_{\rm gas}$ 
in the deeper regions.
It also appears that
the so-called quiet regions have deep-seated magnetic effects.

We find that the near-surface $\beta$ of 
AR$\,9026$ and AR$\,9393$, 
which have largest number of flares among all studied regions,
is anomalously low.
We propose two possible reasons to explain the phenomena.
One is that $P_{\rm gas}$ is larger in the flaring ARs because, as
some theories have proposed,  flares might be caused by 
the larger amount of flux coming from the convection zone.
Another possibility is that 
the relation between $\delta \beta$ and 
$\delta \Gamma_1/\Gamma_1$ derived from our models may not be suitable
for the shallow layers of such highly dynamic and eruptive ARs,
where the variations in $\Gamma_1$ can now come from many sources
rather than solely from the magnetic fields as in our models.
In either case, the inferred $\delta \beta$ may not reflect the actual value.

Our attempt to reproduce the active-region profiles of 
$\delta \Gamma_1/\Gamma_1$ and
$\delta c_T^2/c_T^2$ suggests that
the magnetic fields must be both strong and concentrated in order to produce
the positive $\delta c_T^2/c_T^2$ that is seen in the deeper layers.
Additionally, the
large, positive $\delta \Gamma_1/\Gamma_1$
commonly seen between $0.97R_\odot$ and $0.985R_\odot$
can only be caused by a negative $\delta \beta$ in that region. 
The shifting of ionization zones does not produce matching magnitudes
and locations of the positive $\delta \Gamma_1/\Gamma_1$.

\section*{Acknowledgments}
We wish to thank the referee for detailed comments on this paper.
CHL wishes to thank Dr. Shaun Bloomfield, Prof. Douglas Gough,
and Ms. Claire Raftery
for helpful discussion.
This work utilizes data from the Solar Oscillations
Investigation / Michelson Doppler Imager (SOI/MDI) on the 
{\it Solar and Heliospheric Observatory (SOHO)}. 
SOHO is a project of
international cooperation between ESA and NASA.
MDI is supported by NASA grant NAG5-8878
to Stanford University.
This work is partially supported by NSF grants ATM 0348837 and ATM 0737770
as well as 
NASA grant NNG06D13C to SB.
CHL is also supported by an ESA/PRODEX grant administered by 
Enterprise Ireland.

\end{article}

\begin{thebibliography}{}


\bibitem[Alexander \& Ferguson(1994)]{af1994}
Alexander, D.R., Ferguson, J.W.: 1994,
{\it Astrophys. J.} {\bf 437}, 879.,

\bibitem[Antia \& Basu(1994)]{ab94}
Antia, H.M., Basu, S.: 1994, {\it Astron. Astrophys. Supp.} {\bf 107}, 421.

\bibitem[Bahcall \& Pinsonneault(1992)]{bp1992}
Bahcall, J.M., Pinsonneault, M.H.: 1992,
{\it Rev. Mod. Phys.} {\bf 64}, 885.

\bibitem[Basu et al.(2004)]{Basu_etal2004ApJ610.1157B}
Basu, S., Antia, H.M., Bogart, R.S.: 2004, {\it Astrophys. J.} {\bf 610}, 1157.

\bibitem[Basu et al.(2007)]{BAB2007ApJ654.1146B}
Basu, S., Antia, H.M., Bogart, R.S.: 2007, {\it Astrophys. J.}, {\bf 654}, 1146.


\bibitem[Demarque et al.(2007)]{dem07}
Demarque, P., Guenther, D.B., Li, L.H., Mazumdar, A., Straka, C.W.:
2007, {\it Astrophys. Space Sci.},  {\bf 316}, 31.
\url{http://www.springerlink.com/content/5025w516123252q4/}

\bibitem[Duvall et al.(1993)]{duvall_etal1993}
Duvall, T.L.~Jr., Jefferies, S.M., Harvey, J.W., Pomerantz, M.A.,
1993, {\it Nature} {\bf 362}, 430.

\bibitem[Dziembowski et al.(1990)]{DPS1990MNRAS244.542D}
Dziembowski, W.A., Pamyatnykh, A.A., Sienkiewicz, R.: 1990, {\it Mon. Not.
Roy. Astron. Soc.} {\bf 244}, 542.

\bibitem[Haber(2004)]{Haber2004ESASP559.676H}
Haber, D.A.: 2004, In:
Danesy, D. (ed.),
{\it SOHO 14 Helio- and Asteroseismology: Towards a Golden Future},
{\bf SP-559}, ESA, Noordwijk, 676.

\bibitem[Hill(1988)]{hill1988}
Hill, F.: 1988,
{\it Astrophys. J.} {\bf 333}, 996.

\bibitem[Iglesias \& Rogers(1996)]{ir1996}
Iglesias, C.A., Rogers, F.J.: 1996,
{\it Astrophys. J.} {\bf 464}, 943.

\bibitem[Kosovichev \& Zharkova(1995)]{1995soho.2.341K}
Kosovichev, A.G., Zharkova, V.V.: 1995,
In: Hoeksema, J.T., Domingo, V., Fleck, B., Battrick, B. (eds.),
{\it Fourth SOHO Workshop Helioseismology}, {\bf SP-372-2}, ESA, Noordwijk, 341.

\bibitem[Kosovichev \& Zharkova(1998)]{1998IAUS.185.191K}
Kosovichev, A.G., Zharkova, V.V.: 1998,
In: Deubner, F.-L., Christensen-Dalsgaard, J., Kurtz, D. (eds.)
{\it New Eyes to See Inside the Sun and Stars}, IAU Sympos. {\bf 185},
Kluwer, Dordrecht, 191,

\bibitem[Kosovichev et al.(2000)]{2000SoPh192.159K}
Kosovichev, A.G., Duvall, T.L.~Jr., Scherrer, P.H. 2000, {\it Solar Phys.}
{\bf 192}, 159

\bibitem[Kosovichev et al.(2001)]{2001ESASP464.701K}
Kosovichev, A.G., Duvall, T.L.~Jr., Birch, A.C., Gizon, L., Scherrer, P.H.,
Zhao, J.: 2001,
In: Wilson, A., Pall\'e, P.L. (eds.),
{\it SOHO 10/GONG 2000 Workshop: Helio- and Asteroseismology at the Dawn of the Millennium},
{\bf SP-464},
ESA, Noordwijk,
701.

\bibitem[Lin et al.(2006)]{2006ESASP624E.58L}
Lin, C.-H., Li, L.H., Basu, S.: 2006,
In: Fletcher, K., Thompson, M.J. (eds.),
{\it SOHO 18/GONG 2006/HELAS I: Beyond the Spherical Sun},
{\bf SP-624}, ESA, Noordwijk, 58.

\bibitem[Lin, Soon \& Baliunas(2003)]{2003NewAR.47.53L}
Lin, J., Soon, W., Baliunas, S.L.: 2003, {\it New Astron. Rev.}
{\bf 47}, 53.

\bibitem[Li et al.(2003)]{Li_etal2003ApJ}
Li, L.H., Basu, S., Sofia, S., Robinson, F.J., Demarque, P., Guenther, D.B.:
2003, {\it Astrophys. J.} {\bf 591}, 1267.

\bibitem[Li \& Sofia(2001)]{LiSofia2001ApJ}
Li, L.H., Sofia, S.: 2001, {\it Astrophys. J.} {\bf 549}, 1204.

\bibitem[Lydon \& Sofia(1995)]{ls1995}
Lydon, T.J., Sofia, S.: 1995, 
{\it Astrophys. J. Suppl.} {\bf 101}, 357.

\bibitem[Patron et. al(1997)]{patron_etal1997}
Patron, J., Gonzalez Hernandez, I., Chou, D.-Y., Sun, M.-T., Mu, T.-M., 
Loudagh, S., Bala, B., Chou, Y.-P., Lin, C.-H., Huang, I.-J., Jimenez, A., 
Rabello-Soares, M.C., Ai, G., Wang, G.-P., Zirin, H., Marquette, W., 
Nenow, J., Ehgamberdiev, S., Khalikov, S., TON Team:
1997, {\it Astrophys. J.} {\bf 485}, 869.

\bibitem[Pijpers \& Thompson(1992)]{PT1992}
Pijpers, F.P., Thompson, M.J.:
1992, {\it Astron. Astrophys} {\bf 262}, L33.

\bibitem[Pijpers \& Thompson(1994)]{PT1994}
Pijpers, F.P., Thompson, M.J.: 
1994, {\it Astron. Astrophys} {\bf 281}, 231.


\bibitem[Priest \& Forbes(2000)]{2000mrmt.conf.P}
Priest, E., Forbes, T.: 2000,
In: {\it Magnetic Reconnection: MHD Theory and Applications},
Cambridge University Press, Cambridge, 363.

\bibitem[Rabello-Soares et al.(1998)]{rabello-soares1998soho6.505R}
Rabello-Soares, M.C., Basu, S., Christensen-Dalsgaard, J.:
1998, In: Korzennik, S. (ed.) 
{\it Structure and Dynamics of the Interior of the Sun and Sun-like Stars --
SOHO 6/GONG 98}, {\bf SP-418}, ESA, Noordwijk, 505.

\bibitem[Raftery et. al.(2008)]{Raftery_etal2008}
Raftery, C.L., Gallagher, P.T., Milligan, R.O., Klimchuck, J.A.:
2008, {\it Astron. Astrophys},
{\it submitted}

\bibitem[Rogers \& Nayfonov(2002)]{rn2002}
Rogers, F.J., Nayfonov, A.: 2002,
{\it Astrophys. J.} {\bf 659}, 750.

\bibitem[Rhodes et al.(1998)]{Rhodes_etal1998soho}
Rhodes, E.J.~Jr., Reiter, J., Kosovichev, A.G., Schou, J., Scherrer, P.H.:
1998, In: Korzennik, S. (ed.) 
{\it Structure and Dynamics of the Interior of the Sun and Sun-like Stars --
SOHO 6/GONG 98}, {\bf SP-418}, ESA, Noordwijk, 73.

\bibitem[Schou(1999)]{sch1999}
Schou, J.: 1999, {\it Astrophys. J.} {\bf 523}, L181.

\bibitem[Spruit(1981)]{Spruit1981A&A98.155S}
Spruit, H.C.: 1981, {\it Astron. Astrophys} {\bf 98}, 155.

\bibitem[Trujillo Bueno et al.(2004)]{2004Natur430.326T}
Trujillo Bueno, J., Shchukina, N., Asensio Ramos, A.: 2004, {\it Nature}
{\bf 430}, 326.

\bibitem[Zharkova \& Kosovichev(1998)]{1998ESASP.418..661Z}
Zharkova, V.V., Kosovichev, A.G.:
1998, In: Korzennik, S. (ed.) 
{\it Structure and Dynamics of the Interior of the Sun and Sun-like Stars --
SOHO 6/GONG 98}, {\bf SP-418}, ESA, Noordwijk, 661.


\end{thebibliography}
\end{document}